# MES-P: an Emotional Tonal Speech Dataset in Mandarin Chinese with Distal and Proximal Labels

Zhongzhe Xiao, *Member, IEEE*, Ying Chen, Weibei Dou, *Member, IEEE*, Zhi Tao, Liming Chen, *Senior Member, IEEE*

***Abstract*—Emotion shapes all aspects of our interpersonal and intellectual experiences. Its automatic analysis has there-fore many applications, *e.g.*, human-machine interface. In this paper, we propose an emotional tonal speech dataset, namely Mandarin Chinese Emotional Speech Dataset - Portrayed (MES-P), with both distal and proximal labels. In contrast with state of the art emotional speech datasets which are only focused on perceived emotions, the proposed MES-P dataset includes not only perceived emotions with their proximal labels but also intended emotions with distal labels, thereby making it possible to study human emotional intelligence, *i.e*. people emotion expression ability and their skill of understanding emotions, thus explicitly accounting for perception differences between *intended* and *perceived* emotions in speech signals and enabling studies of emotional misunderstandings which often occur in real life. Furthermore, the proposed MES-P dataset also captures a main feature of tonal languages, *i.e.*, tonal variations, and provides recorded emotional speech samples whose tonal variations match the tonal distribution in real life Mandarin Chinese. Besides, the proposed MES-P dataset features emotion intensity variations as well, and includes both *moderate* and *intense* versions of recordings for joy, anger, and sadness in addition to neutral speech. Ratings of the collected speech samples are made in valence-arousal space through continuous coordinate locations, resulting in an emotional distribution pattern in 2D VA space. The consistency between the speakers' emotional intentions and the listeners' perceptions is also studied using Cohen's Kappa coefficients. Finally, we also carry out extensive experiments using a baseline on MES-P for automatic emotion recognition and compare the results with human emotion intelligence.**

## I. INTRODUCTION

EMOTION is one of the most important aspects in human communication and is conveyed in multiple ways, *e.g.*, speech, facial expressions, body gestures, *etc*. Studying machine learning based emotion recognition has become a major research topic in the field of affective computing [1][2]. It can be useful in many potential applications, *e.g.*, multimedia indexing, natural human-computer interaction, or emotion monitoring in extreme working conditions. Multi-modal emotional research [3][4][5][6][7][8] on speech, facial image/video clips [9], and in some cases, physiological signals [10], has achieved remarkable results. However, there are many situations where only speech signals can be conveniently collected, such as conversations via telephone, in locations with no cameras, or poorly illuminated. Thus, emotion research on single modal speech signals is still of great importance.

For machine-based emotion recognition, datasets of emotional speech, with different emotional description approaches and in different languages, form the basic source data in the field. In these emotional speech datasets, collected speech samples are subjectively evaluated by raters to provide emotional labels. Nevertheless, in speech communication, emotions are expressed or encoded into vocal sentences by speakers as **distal** indicators, and perceived or decoded by listeners as **proximal** percepts. Distortions could occur during emotion transition from *distal* end (encoding) to *proximal* end (decoding), which result in emotion misunderstandings. In conventional approaches of emotional speech collection, subjective evaluation is used as a screening stage to ensure the consistency of the collected emotional samples. Speech samples judged as poorly expressing the intended emotion by majority of raters are simply rejected. The screening by rejection makes the current datasets to only reflect the perceived emotions, while ignoring the speakers' emotional intentions. In contrast to state of the art emotional datasets, we argue in this work that, in collection of portrayed emotional speech datasets, the emotional states that the acting speakers intend to express should be taken as **distal** (intentional) emotional labels themselves, and the evaluation by the raters (perceived emotions) should be taken as **proximal** (perceptual) emotional labels. Indeed, the joint distal and proximal label of emotional speech samples make it possible to investigate people's emotional intelligence, *i.e*., assessment of individual speaker's emotion expressivity when proximal labels are used as ground truth or inversely evaluation of individual rater's skill of correctly understanding emotions when distal labels are used as ground truth. In both cases, the disagreement between distal and proximal labels characterizes situations of emotional misunderstanding which unfortunately often occur in real-life communications and constitute source of conflicts. Therefore, none of the collected speech samples should be rejected.

Mandarin Chinese is a tonal language. It is well known that the shape of F0 trace differs in different emotional states [11].

This work is sponsored by Basic Research Program of Jiangsu Province (Natural Science Foundation), China, No. BK20140354, and in part by the French Research Agency, l'Agence Nationale de Recherche (ANR), through the project Jemime under grant ANR-13-CORD-0004-02.

Zhongzhe Xiao, Ying Chen, and Zhi Tao are with School of Optoelectronic Science and Engineering, Soochow University, Suzhou, 215006, P. R. China. (e-mail: xiaozhongzhe@suda.edu.cn, taoz@suda.edu.cn).

Weibei Dou is with the Tsinghua National Laboratory for Information Science and Technology, Department of Electronic Engineering of Tsinghua University, Beijing, 100084, P. R. China, (e-mail: douwb@tsinghua.edu.cn)

Liming Chen is with the Universit éde Lyon, Centre National de la Recherche Scientifique, Ecole Centrale de Lyon, LIRIS, UMR5205, F-69134, France, (e-mail: liming.chen@ec-lyon.fr); He is also with Beijing Advanced Innovation Center for Big Data and Brain Computing (BDBC), Beihang University, China

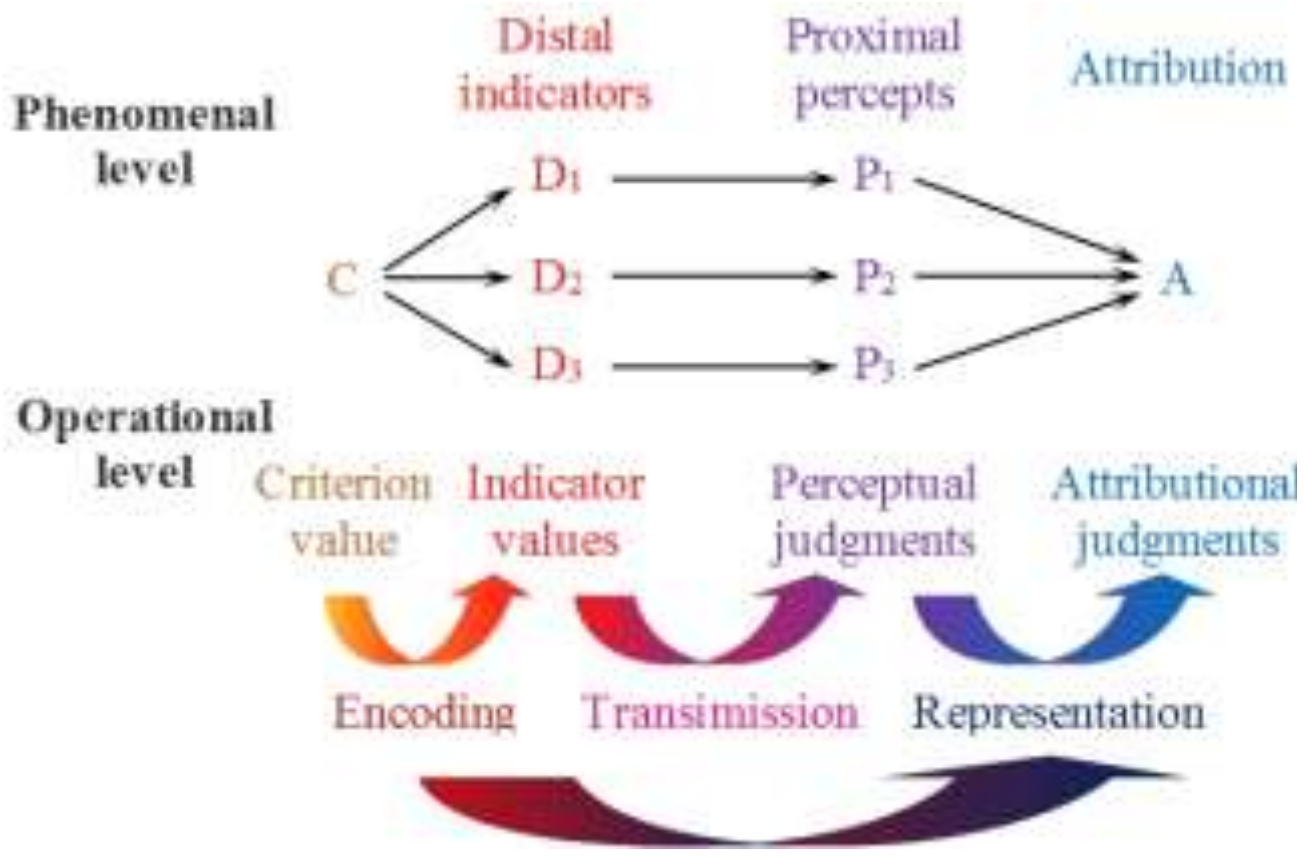

Fig. 1. Brunswikian lens model of the vocal communication of emotion

The tones in Mandarin Chinese correspond to specific shapes of F0 traces. The basic F0 trace shapes and their deformations according to emotion changes, which are ignored in existing Mandarin emotional speech datasets, should also be considered. Moreover, in real-life speech communications, emotions are expressed through different intensities with the domination of neutral and moderate emotions in vocal communications.

In this work, we propose an emotional speech dataset in Mandarin Chinese, namely MES-P (Mandarin Chinese Emotional Speech Dataset - Portrayed). The dataset comprises emotion intensity, and includes 7 emotional states as moderate and intense versions of joy, anger, and sadness as well as neutral speech. In the recording process of MES-P, speakers were asked to act the 7 emotional states by reading sentences of predefined verbal scripts. The scripts were designed to roughly follow the tonal distribution of the oral Mandarin in real life, and cover all types of vowels and consonants in Mandarin. As *distal* labels, the speakers only save speech samples that they think faithfully convey their emotional intentions. In the evaluation stage, the raters do not specify discrete emotional states but assign locations for the speech samples in a dimensional valence/arousal space. They are used as *proximal* labels.

The contributions of the proposed MES-P dataset are three-fold:

- MES-P is the first emotional speech dataset which record both the *distal* and *proximal* labels of emotional speech. It can be used for studies of emotion intelligence, *e.g.*, emotional misunderstanding in speech communications.
- MES-P is also the first emotional speech dataset in Mandarin Chinese which follows the tonal distribution in oral speech. The influence of tones and emotions to the intonation could be analyzed based on MES-P, as a complement to the existing datasets.
- We also provide a baseline performance for emotion classification using MES-P and compare the baseline results with human skills of emotion understanding.

Section II introduces emotion description approaches and existing datasets. Section III describes in detail the process for building the proposed MES-P dataset, including script design, data collection, and emotion rating. Section IV studies the effectiveness and consistency of the proposed MES-P dataset. Section V analyses human emotion intelligence using distal and proximal labels. Section VI provides a baseline performance on MES-P in comparison with human emotion recognition skills. Section VII concludes this work and give some hints on our future research.

## II. Emotion Taxonomy and Existing Datasets

### A. Emotion Description in Vocal Communication

Emotion transmitted by the human voice, or the vocal communication of emotions as defined by Scherer [12], is present in two stages: the speaker's expressional states (*distal* indicators), or the listener's perceptual states (*proximal* percepts). They are described by the Brunswikian lens model [13], as shown in Fig. 1. Objective measurements are needed to capture the subjective emotions conveyed in speech signals. According to Brunswik's terminology, starting from the observer's or the listener's viewpoint, the acoustic changes of speakers can be seen as *distal* cues (distant from the observers), which include emotion specific patterns caused by changes in respiration, intonation and articulations correlated with the speaker's underlying ecological validity. The *distal* cues are transmitted along with speech signals and perceived by the listener's auditory perceptual system, as *proximal* cues (close to the observers), with some uncertainty in definition [14], in the emotional vocal communications. The proximal cues are based on the distal cues, which may be modified or distorted during transmission or in perception [12]. Therefore, there might exist divergence between *intended* emotions expressed by speakers and *perceived* emotions by listeners.

Emotional speech datasets are fundamental for studies of vocal emotions. According to Scherer [9], such resources can be classified into three categories, *i.e*., natural vocal expressions, induced emotional expressions, and simulated emotional expressions. Simulated, or "portrayed" vocal emotions have been the preferred way for obtaining emotional voice samples in state of the art. Actors are asked to express specified emotions, usually through a given content. This is the way to get the most intense, typical expressions under highly controlled conditions; however, actors can overemphasize some obvious cues and miss some subtle ones. As such, emotions conveyed in naturally occurring voices have the highest ecological validity. However, it has been shown that in real life, the emotion expressed in voice is relatively dispersed and is dominated by neutral speech or moderate emotions [15]; thus, speech samples with typical intense emotions are rare. This type of voice sample is generally collected in a poor recording environment, and it is difficult to determine the nature of the underlying emotion.

The major descriptions in emotion taxonomy include discrete emotional states and dimensional emotion spaces. Most typical discrete descriptions of basic emotions originate from Ekman's big six [16][17]; more discrete emotions have been considered in previous studies aiming at various applications. The dimensional approach dates back to Wundt [18] in 1874 and maps emotional states into a two or three-dimensional space. A widely used emotional space is two dimensional, including a valence dimension (pleasant - unpleasant, agreeable - disagreeable), and an arousal dimension (active - passive) [19]. This two dimensional emotional space is abbreviated as VA space in the subsequent sections.

TABLE I
EMOTIONS IN THE DATABASES AS SUMMARIZED BY VERVERIDIS [25]

| Emotion | Number of databases |
|---|---|
| Anger | 26 |
| Sadness | 23 |
| Happiness | 13 |
| Fear | 13 |
| Disgust | 10 |
| Joy | 9 |
| Surprise | 6 |
| Boredom | 5 |
| Stress | 3 |
| Contempt | 2 |
| Dissatisfaction | 2 |
| Shame, pride, worry, startle, elation, despair, humor,… | 1 |

### B. *Existing Datasets*

Datasets are essential resources in machine learning. Multiple emotional datasets or corpora in various forms have been collected, including datasets with only speech, or speech together with human facial information, and across a wide range of languages.

Representative early datasets, which came into use before the year 2000, include SUSAS, which focused on stress [20], and the widely spread Danish Emotional Speech database DES [21]. A number of other datasets emerged near 2000 [22][23], including the famous public database in German, Emo-DB [24], which is still used in many works as baseline data. A summary was made by Ververidis [25] who compared 32 early emotional speech datasets, including 21 simulated /portrayed datasets, 8 natural / spontaneous datasets, and 3 semi-natural datasets, in more than 10 languages (English, German, Danish, Chinese, *etc*.). These datasets were com-piled for various purposes in emotion research, *e.g.*, emotion recognition, emotion speech synthesis, emotion perception by humans, *etc*. The frequency of emotion occurrences in these 32 datasets is summarized in Table I. The most common recordings are anger, sadness, happiness, fear, disgust, joy, surprise, boredom. Similar conclusions were drawn in [26]. While the joy emotion is mostly combined with happiness due to their similarity, anger, sadness, and joy were the three most commonly labeled emotions in the early datasets.

A more recent database, and one of the most important large scale spontaneous emotional speech databases, was developed by Steidl [27]. This database was used as a speech resource in the challenge of emotional speech at INTERSPEECH 2009 [28]. It is special in that the speakers were children and the emotions were induced by interactions between the children and a speech robot. This database has been used by a number of researchers after the challenge. New emotional speech datasets are continuously appearing [29][30][31][32].

With an increasing number of emotional speech datasets in different languages, studies across datasets/languages have become one of the major topics in vocal emotion analysis [33][34][35][36][37][38]. These works showed that emotions in speech could be recognized across different languages but that the performances were generally not as good as those in tasks with only one language. Thus, the language property is also a key factor in vocal emotion analysis. As a tonal language, there are a number of distinct characters in Mandarin Chinese worth studying in emotion research. There are also several emotional speech datasets in the Chinese language [10][39], aiming at different application backgrounds. In our work, we are interested in the different presentations through different tones in emotional speech, the consistency and difference between the emotion expression intention of the speakers and the emotion perception of the listeners, and the distribution of the typical discrete emotional states in the VA space. To address these issues, we developed a new Mandarin Chinese emotional speech dataset with portrayed emotions. The next section describes the development process of the dataset.

## III. SPEECH DATA COLLECTION

The major goal of our dataset is to enable investigation of intended emotions through distal cues, and perceived emotions via proximal ones as well as their relationships in daily life oral communication. As a result, we have collected emotional speech samples from nonprofessional actors instead of professional actors who have tendency to exaggerate emotions. However, in order to ensure the reliability of expressed emotion intentions as distal labels by these nonprofessional actors, our dataset is designed as a portrayed one in a highly controlled way. The speech samples were collected by asking volunteers to act out the required 7 emotional states using a script with 16 sentences in Mandarin Chinese. The collected speech samples were then evaluated by 7 raters as coordinate locations in the VA space.

### A. *Selection of Emotions*

As summarized in Section II-B, anger, sadness, and joy/happiness were the most commonly occurred typical emotions in the state of the art datasets. They are also relatively common in acting. In addition, neutral speech or speech with no emotional tendency, also needs to be considered as a base state for emotional speech. In this work, these frequently occurred emotions in state of the art emotional speech datasets, *i.e.*, neutral, joy/happiness, anger, and sadness, are also selected in our dataset. Such a selection also makes it possible to the studies on emotion recognition across datasets and languages. However, given the domination of moderate emotions in real-life [15] and to stick tight with real-life oral communication, we further introduce emotion intensities in our dataset. Specifically, we define two different emotional states which could be

TABLE III
EMOTIONAL STATES AND INDEXES

| Index | Emotional states | |
|---|---|---|
| 1 | Neutral | |
| 2 | Joy/Happiness | Moderate/Slight |
| 3 | | Intense/Strong |
| 4 | Anger | Moderate/Slight |
| 5 | | Intense/Strong |
| 6 | Sadness | Moderate/Slight |
| 7 | | Intense/Strong |

TABLE II
VERBAL SCRIPT IN EMOTIONAL SPEECH COLLECTION

| Index | Verbal Script (in Pin Yin)<br>English Translation |
|---|---|
| 1 | Wǒ kàn jiàn zhuō zi shàng yǒu yí gè fěn sè de xī hóng shì.<br>I saw a pink tomato on the table. |
| 2 | Mì fēng fēi qǐ lái huì yǒu wēng wēng de shēng yīn.<br>The bees will have a buzzing sound when flying. |
| 3 | Tā měi gè zhōu mò dōu huì dào nàr* q ù y ít àng.<br>She goes there every weekend. |
| 4 | Zài guò jǐ tiān jiù yào kāi shǐ luò yè le.<br>The leaves will begin to fall in a few days. |
| 5 | Tā men zhù de shì liù ge rén yì jiān wū zi de sù shè.<br>They live in a dorm room for six people. |
| 6 | Ér qiě, qù nián qī yuè tā qīn zì xiě le yì shǒu gē.<br>And, he wrote a song himself last July. |
| 7 | Kě shì zhè suǒ xué xiào yǒu jǐ bǎi wèi lǎo shī.<br>But there are hundreds of teachers in this school. |
| 8 | Jīn nián de méi yǔ jì shí zài shì tǐng cháng de.<br>The rainy season this year is really long. |
| 9 | Tā hěn xī huān yòng gāng qín tán huān lè sòng.<br>He likes to play “Ode to Joy” with piano. |
| 10 | Xiàn zài yǐ jīng hěn shǎo yǒu rén yòng cí dài lù yīn jī le.<br>Few people use tape recorders now. |
| 11 | Qián tiān shí táng gēn běn jiù méi yǒu mài bāo zi.<br>No steamed buns were sold in the canteen the day before yesterday. |
| 12 | Xiǎo Lán jīng cháng pàn zhe kuài diǎn xià xuě.<br>Xiao Lan always hopes to snow earlier. |
| 13 | Tā zuì xǐ huān chàng de gē shì zhù nǐ shēng rì kuài lè.<br>His favorite song is “Happy birthday to you.” |
| 14 | Nǐ měi cì qù yóu yǒng dōu bú ài tú fáng shài shuāng.<br>You don't like to apply sunscreen every time you go swimming. |
| 15 | Du ìle, zu ótiān dǎ léi de shí hòu mén qián yǒu yì zhī hǎi ōu.<br>By the way, there was a seagull in front of the door yesterday. |
| 16 | Yán sè yīng gāi tú de gèng jūn yún diǎn jiù hǎo le.<br>The color should be painted more evenly. |

*: A rhotic accent written with two characters but pronounced as one syllable

described by the same or similar words, such as “anger” or “joy”, as an emotion family, and collect a moderate version and an intense version for each emotion family. Together with neutral, the total number of emotional states rises to 7 in our dataset. This also makes it possible for future expansion of our dataset from discrete emotional states to continuous emotions. Table II lists these seven emotional states. They are indexed from 1 to 7 and serve as discrete distal labels by the speakers when recording the speech samples.

While the selected seven emotional states, i.e., neutral state, three moderate emotions, and three intense emotions, are still limited in comparison with the full range of emotions which can be expressed by a human voice in daily oral communications, increasing the number of emotions proves to be very difficult for speakers, especially for no professional actors, to comprehend and demonstrate the assigned emotions and could lead to poor quality of emotions expressed in speech. We find that the selection of these 7 emotional states is a reasonable trade-off between the reliability of collected emotional speech samples and the coverage of emotions.

### *B. Setting of verbal script*

Sixteen sentences are designed as verbal scripts to be acted in the seven chosen emotional states by the speakers in collecting the dataset.

In order to present the pronunciations of the script, Pinyin system, which allows the notation of tones of each syllable, is applied in this paper. Pinyin, literally means “spelled sounds”, is the official romanization system for Standard Chinese. Based on several earlier forms of romanization of Chinese language, Pinyin system was developed in the 1950's by many linguists [40]. It was published in 1958 by the Chinese government with several later revisions [41]. It became an international standard of ISO in 1982 [42].

The sentences (in Pinyin) and their English translations are listed in Table III. All sixteen sentences are emotionally neutral in meaning. In total, there are 207 syllables in the sixteen sentences. The script is designed to cover all tones and all types of vowels and consonants in Mandarin Chinese.

*1) Tones:* Different emotions may cause significant changes in speech pitch contour, both in terms of mean value and gradient. For example, there could be quick pitch drops in speech with anger emotion [43]. As a tonal language, there are four main tones and a light tone in Mandarin Chinese. Each tone presents an inherent shape of pitch contour; thus, the influence of tones needs to be considered in the study of speech emotion.

The four main tones are high tone, mid rising tone, low dipping tone, and high falling tone. They are indexed using the numbers 1 to 4. In the Pinyin system, the tones are also assigned with symbols imitating the shapes of pitch contour that are placed above the key vowels in syllables. There is a $5^{th}$ tone, a light tone, which normally appears as auxiliary words and is pronounced with less intensity and in shorter syllables, and has no extra tone symbol in Pinyin. The pitch level of the light tone usually depends on the preceding syllable. Chao [44] proposed a tone letter system early in 1930 that presents pitch contour in five levels. The numeric levels and symbols in Pinyin of the tones are listed in Table IV.

Fig. 2 illustrates the F0 (Fundamental frequency, which can be perceived as pitch) contour of the vowel "a" in each tone. The F0 contours for the four main tones and the light tone are presented in Fig. 2 (a). The light tone here is pronounced after a syllable with the 4$^{th}$ tone (high falling tone); thus, it holds a low pitch level as an extension to the end of the previous syllable. The F0 contours are normalized to syllable duration, and plotted together in Fig. 2 (b), which shows great similarity with Chao's description in Fig. 2 (c). One of the subjects to be studied with our emotional speech dataset will be the influence on pitch contour by emotion for different tones.

TABLE IV
TONES IN CHINESE MANDARIN

| Tone Number | Tone Name | Chao's tone numerals | Pinyin for vowel "a" |
|---|---|---|---|
| 1 | High tone | 55 | ā |
| 2 | Mid rising tone | 35 | á |
| 3 | Low dipping tone | 214 | ǎ |
| 4 | High falling tone | 51 | à |
| 0 | Light tone | -- | a |

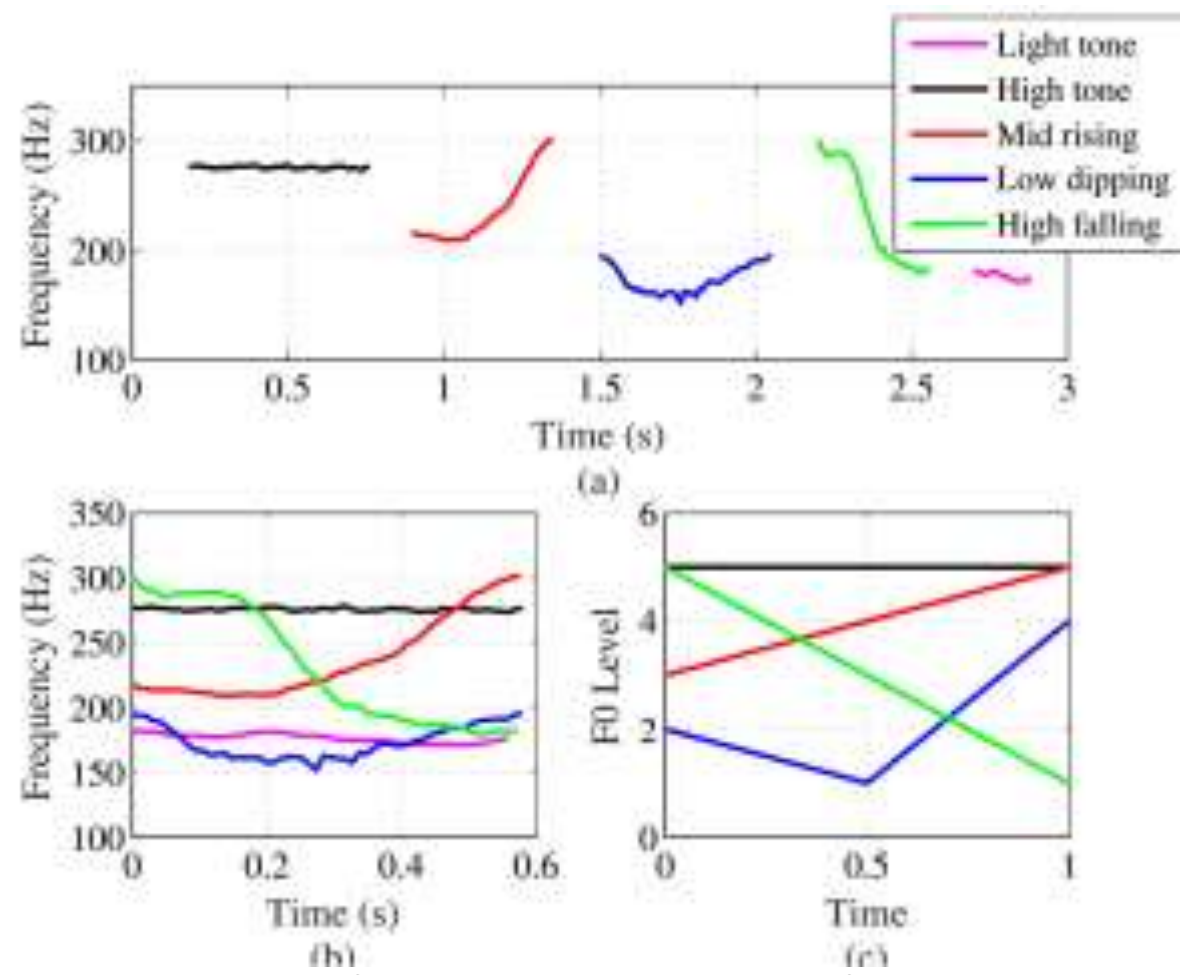

Fig. 2. Fundamental frequency contour in tones of Chinese Mandarin (a) F0 contours for the vowels ā, á ǎ, à a; (b) F0 contours normalized to syllable duration; (c) Relative F0 level according to Chao's tone description.

The distribution of the tones of the syllables in the 16 sentences of this dataset is analyzed to ensure it follows the real usage situation in spoken Mandarin. On the verbal script (in Pinyin), we marked syllables with different colors for different tones in Fig. 3. Most sentences (13 out of 16) cover all four main tones. The light tone also appears in 12 of the sentences. The numbers for each tone in the whole script are listed in Table V, where the light tone is indexed as tone 0.

To evaluate the effectiveness of the script setting in terms of tones, the overall tone distribution in Mandarin Chinese is used as a baseline. There are several different versions of character frequency statistics tables for simplified Chinese. A character frequency table [45][46] accessible on the Internet is chosen in calculation of this baseline, which is based on the 6763 characters in the GB2312 standard [46] that covers over 99 % of the characters of contemporary usage [47]. The statistics on tone distributions are also listed in Table V. More than 30% of the characters are in the 4$^{th}$ tone, approximately 25% in the 1$^{st}$ and 2$^{nd}$ tones, and less than 20% in the 3$^{rd}$ tone. When the frequency of character usage is considered, the proportion of the 3$^{rd}$ tone is slightly increased over its usage in characters, while the proportions of the 1$^{st}$ and 2$^{nd}$ tones drop. There are only a very small number of characters in Mandarin Chinese in the light tone, but they are frequently used as auxiliary words, thus the light tone's proportion in pronunciation is obviously higher than its character proportion. A comparison of the difference between the distribution of tones in the script of our dataset and the baseline is shown in Fig. 4. The two curves fit well, especially for the 1$^{st}$, 3$^{rd}$, and 4$^{th}$ tones, which shows that the distribution of tones in our dataset is a reasonable reflection of the real-life usage of spoken Mandarin, thus demonstrating the effectiveness of the tone setting in the script used for our emotional speech dataset.

TABLE V
DISTRIBUTION OF TONES

| Tones | Character distribution | Pronunciation distribution | Occurrence in script | Distribution in script |
|---|---|---|---|---|
| 1 | 25.01% | 22.85% | 46 | 22.22% |
| 2 | 25.92% | 19.73% | 35 | 16.91% |
| 3 | 17.12% | 18.73% | 38 | 18.36% |
| 4 | 31.58% | 31.42% | 68 | 32.85% |
| 0 | 0.35% | 7.25% | 20 | 9.66% |

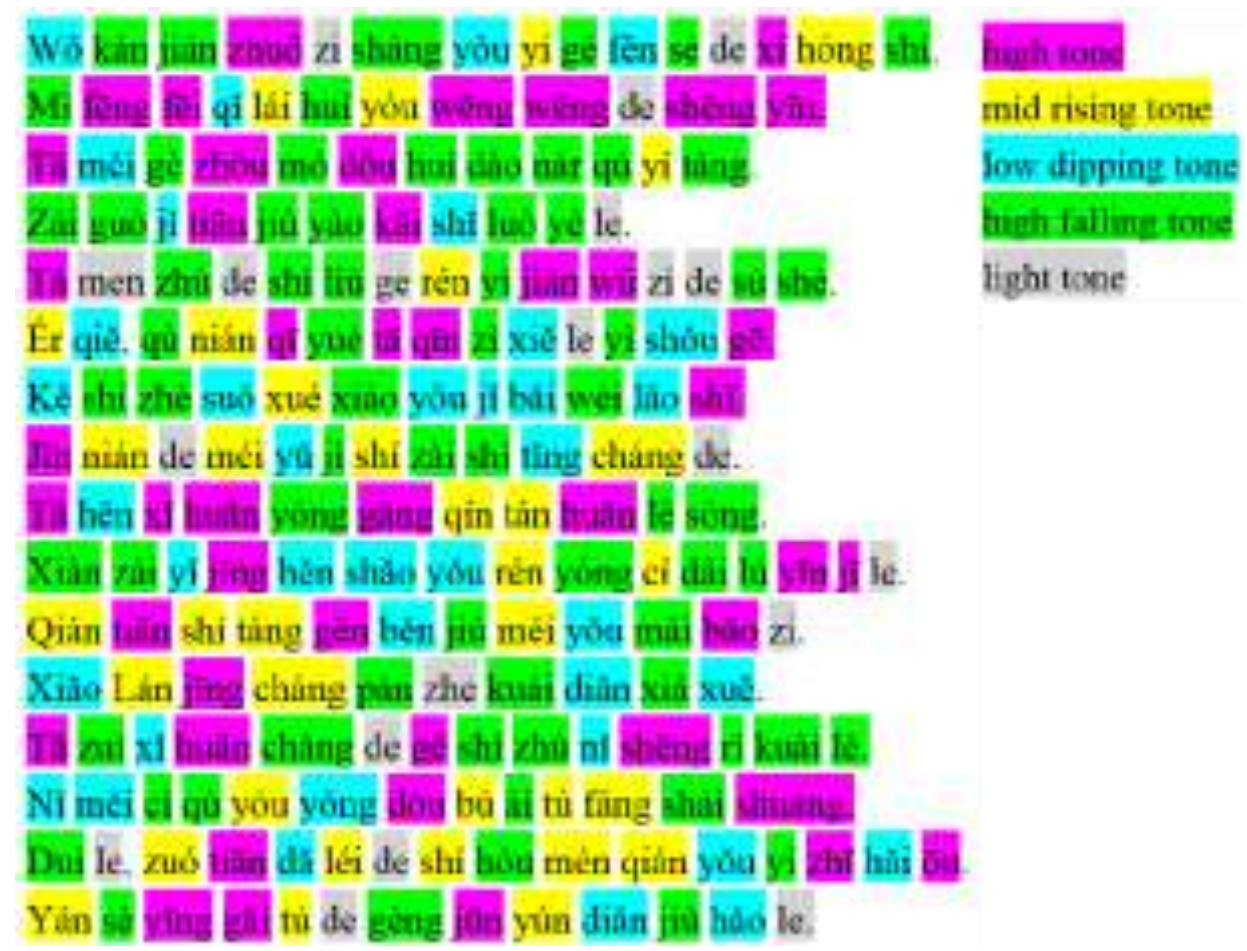

Fig. 3. Tones in the verbal script of our dataset.

*2) Vowels*: The vowels in speech are essential in conveying emotions. Emotional states cause variances in the states of vocal cords and the vocal tract and involve their shapes and tension when producing speech. These variances are reflected by basic speech parameters, such as fundamental frequency and formants, which are present in vowels. The vowels in Mandarin Chinese, which are also called "finals" in the Pinyin system, are classified as simple vowels and compound vowels including both normal compound vowels and nasal vowels. The vowels are listed in Table VI by type. Due to the scale of our dataset,

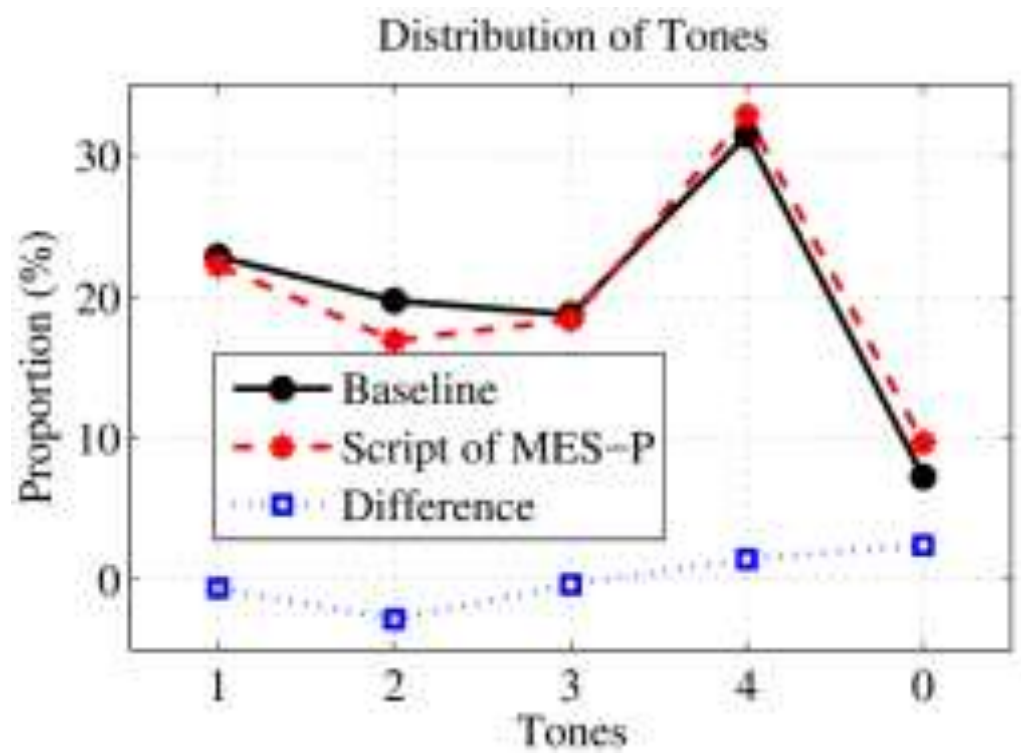

Fig. 4. Comparison of tones distribution, baseline vs. MES-P script.

TABLE VI
VOWELS IN CHINESE MANDARIN

| Types | Vowels |
|---|---|
| Simple vowels | a, o, e, ê, i, u, ü[$], er, -i (front)[#], -i (back)[+] |
| Normal compound vowels | ai, ei, ao, ou, ia, ie, ua, uo, üe, iao, iou, uai, uei |
| Front nasal vowels | an, en, in, ün, ian, uan, üan, uen |
| Back nasal vowels | ang, eng, ing, ong, iang, uang, ueng, iong |

[*]: Only as modal word when used in isolation, does not follow any consonant. Used in compound vowels ie, üe

[$]: Written as u after consonants j, q, x, y

[#]: Vowel as extension of consonants z, c, s

[+]: Vowel as extension of consonants zh, ch, sh, r

there are only 207 syllables in the script. We do not expect to follow the natural distribution of vowels in real spoken situations. Instead, we ensure that all the simple vowels and all types of compound vowels are covered. The individual simple vowels and compound vowels types are denoted using different colors in Fig. 5(a) and Fig. 5(b). All 16 sentences in the script contain both simple vowels and compound vowels. In total, there are 86 simple vowels, 61 normal compound vowels, 36 front nasal vowels, and 24 back nasal vowels in our script. The front and back nasal vowels can be analyzed in two ways, separately or together, because some people (native speakers of the Chinese language) in some areas tend to confuse them.

3) Consonants: The consonants in Mandarin Chinese are also called "initials" in the Pinyin system. The analysis of consonants is not a common subject in the study of emotional speech. As the shape and position of lips and tongue also vary due to changing emotion and may introduce airflow obstructions, we also propose to cover all the consonants in Mandarin for possible analysis. According to their airflow obstruction, the consonants are listed in Table VII, and denoted using different colors in Fig. 6.

## C. Speakers in recording

16 volunteers (8 female speakers and 8 male speakers) participated in the collection of this dataset. All the speakers were university students (undergraduate & graduate), aged from 20 to 23 years old. None of the speakers were professional actors, because a major purpose in collecting this emotional speech dataset is to evaluate the skill of understanding and expressing emotions by ordinary people.

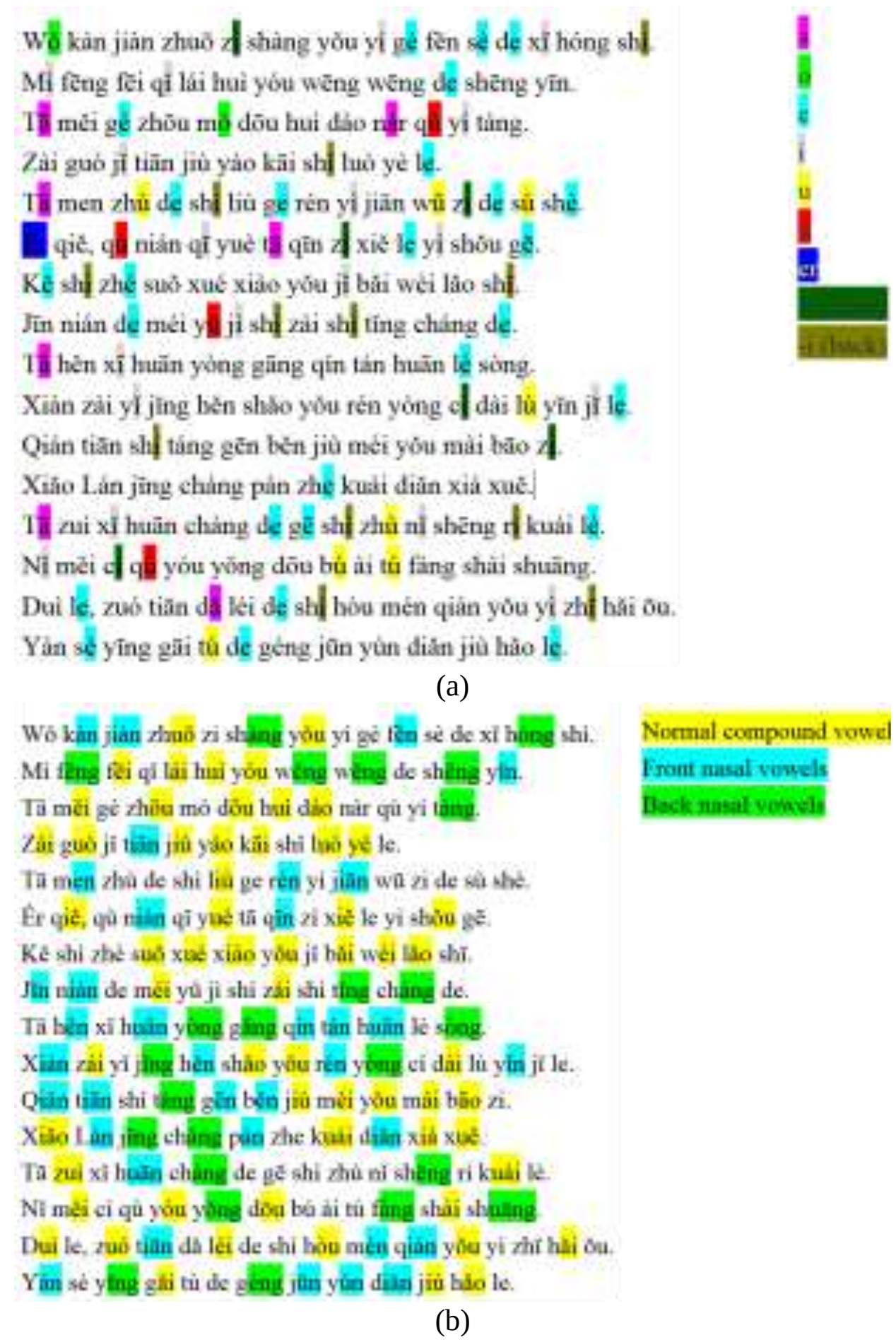

Fig. 5. Vowels in the script (a) Simple vowels (b) Compound vowels

TABLE VII
CONSONANTS IN CHINESE MANDARIN

| Types | Consonants |
|---|---|
| Plosive | b, p, d, t, g, k |
| Affricates | z, c, zh, ch, j, q |
| Fricative | f, h, s, sh, r, x |
| Nasal | m, n |
| Lateral | l |
| Zero initials[*] | y, w |

[*]: Replace i/ ü/u at the beginning of syllables without consonant

In order to ensure the quality of the emotional recordings and the reliability of their annotations in the MES-P dataset, speakers were carefully selected based on several criteria. The 1st criterion is that all the speakers are native speakers of Mandarin Chinese, able to speak Mandarin naturally, fluently, and easily intelligible. The 2$^{nd}$ one is that none of the speakers had any vocal diseases. However, because there are many different dialects in the Chinese language, quite a large number of native speakers speak Mandarin with accents from different areas in real-life situations. Thus, a slight accent variation was also allowed in the selection of speakers during the collection of our dataset so that the collected speech samples be consistent with the actual situation.

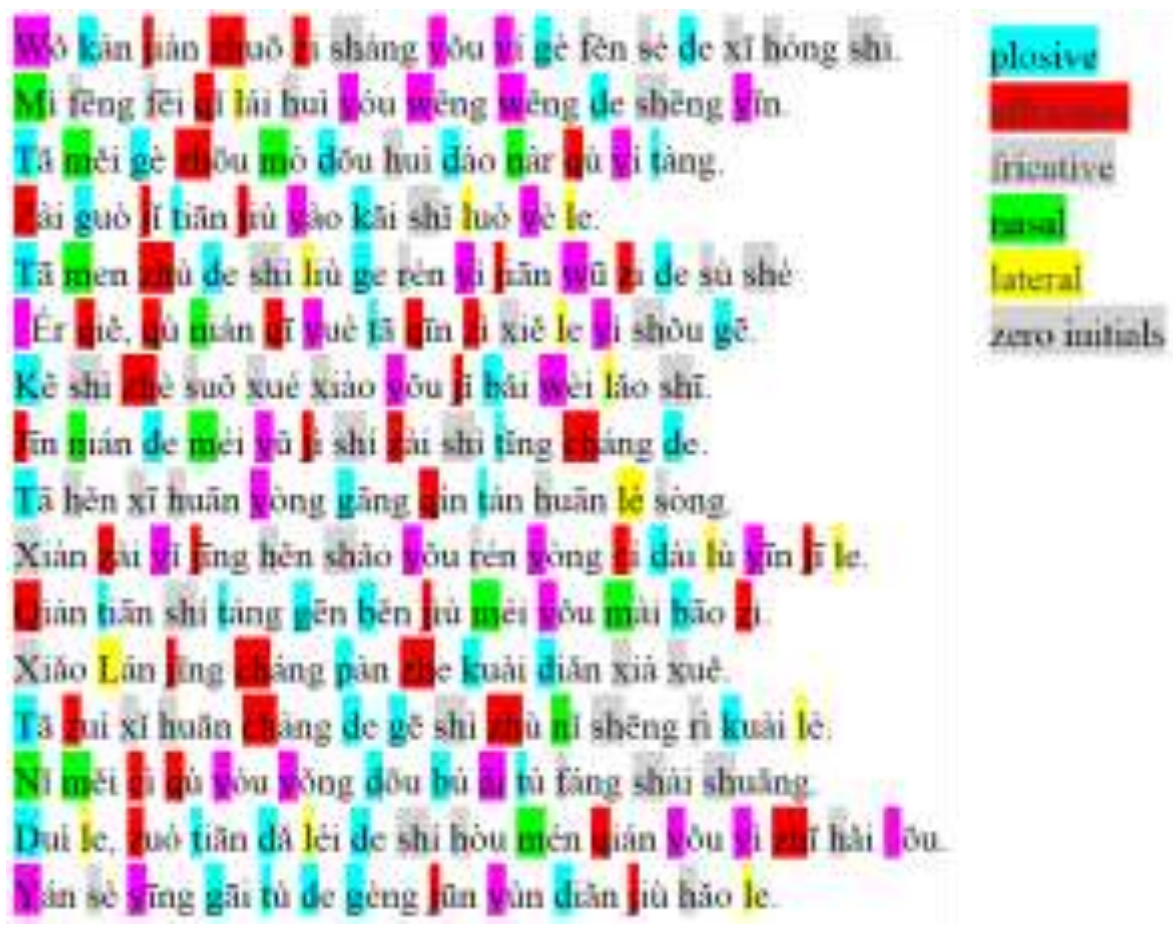

Fig. 6. Consonants in the script.

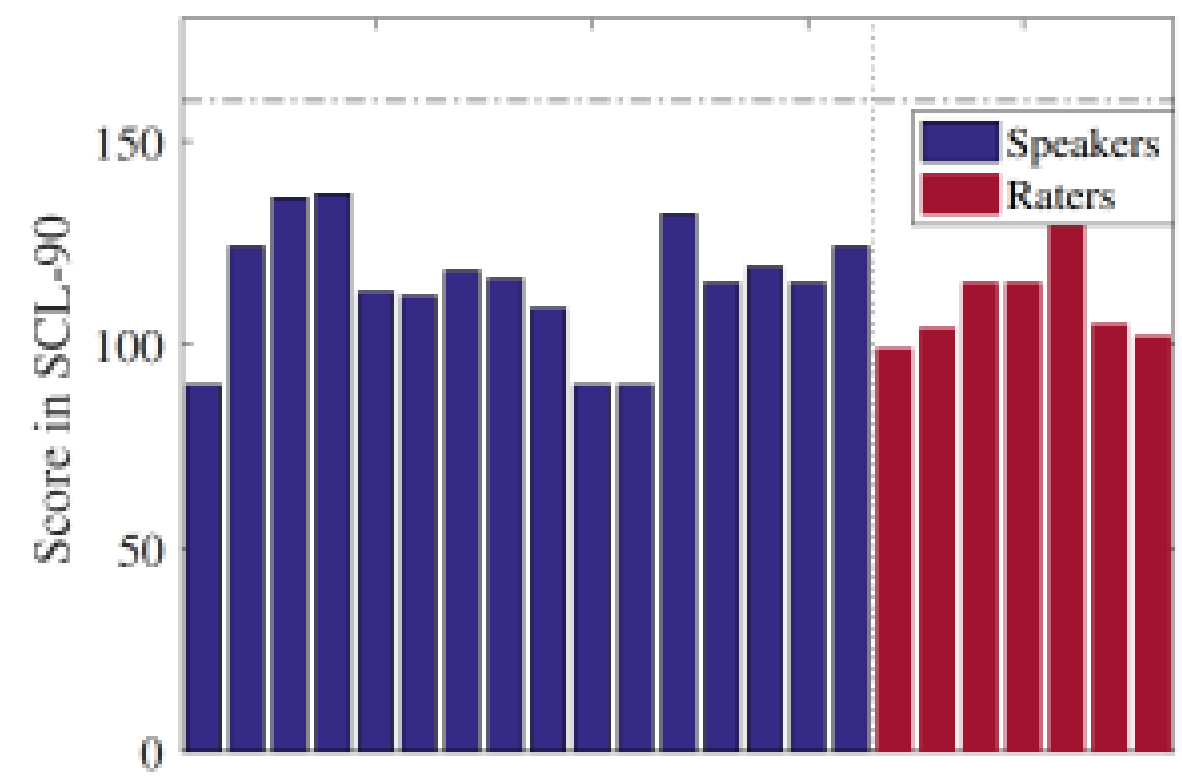

Fig. 7. Total SCL-90 scores for the speakers and raters.

TABLE VIII
THE FACTORS IN SCL-90

| Index | Factor | Index | Factor |
|---|---|---|---|
| 1 | Somatization | 6 | Hostility |
| 2 | Forced Symptoms | 7 | Photic anxiety |
| 3 | Interpersonal sensitivity | 8 | Paranoia |
| 4 | Depression | 9 | Psychoticism |
| 5 | Anxiety | 10 | Other |

The 3$^{rd}$ criterion ensures that all the speakers have nor-mal ability in expression and perception of emotion so that the proposed dataset has no bias with emotions expressed in daily interpersonal communication. For this purpose, all the speakers were requested to fill the SCL-90 [48] survey form, which is a widely used self-assessment scale. SCL-90 (Symptom Checklist) is a self-report questionnaire, which was originally designed towards symptomatic behavior of psychiatric outpatients, and has been developed into a psychiatric case-finding instrument, and a descriptive measure of psychopathology [49][50].

Specifically, the speakers and raters were asked to answer the 90 questions in SCL-90 with scores 1 to 5, where 1 refers to “Not at all”, 2 “A little bit”, 3 “Moderately”, 4 “Quite a bit”, and 5 “Extremely”. The total score ranges from 90 to 450. Scores lower than 160 indicate normal people without obvious psychiatric symptoms. The scores for the speakers and raters participating to the collection of this dataset are illustrated in Fig. 7, where blue bars designate the speakers and brown bars correspond to the raters. The upper limit score for normal psychological healthy people, *i.e.*,160, is marked in dash-dotted line in Fig. 7 as reference. The highest score is 137, which is still far below the reference line, thereby suggesting that the psychological behaviors of all the speakers and raters are distributed within the normal range.

The 90 questions in SCL-90 are also grouped to reflect 10 different psychological factors, as listed in Table. VIII. Factors 2 to 8 are more intensively related to ability of emotion expression and perception. The scores for the speakers and raters in each of the 10 factors are illustrated in Fig. 8. The indexes of the factors are marked as horizontal axis labels. The background colors in this figure correspond to the symptomatic degrees, where yellow areas range from the lower limit to low symptomatic, blue areas range in medium symptom, and green areas range in high (obvious) symptom. As can be seen in Fig. 8, all the scores fall in the range of low symptomatic areas, except two cases which reach the bottom of medium symptomatic area of factor 2 (Forced Symptoms / Obsessive-compulsive). However, given the overall scores from the SCL-90, far below the sane score of 160, we can assume reasonably that all the speakers and raters are affectively normal, and capable of recoding as well as interpreting the speech samples for our emotional speech dataset without notable bias.

### D. *Recording Process*

The speech samples were recorded in a quiet classroom. To minimize the influence of environmental noise, the recording was performed on weekends when few people were present. The equipment used for recording included a notebook computer Dell Inspiration N4110, a monitoring headphone Audio Technica AHT-SR5, and a cardioid condenser microphone Audio Technica AT2020USB. The characteristics of this microphone are shown in Fig. 9 [51].

The speech sample collection process was divided into a preparation session and a recording session.

In the preparation session, a staff member first explained the purpose of the data collection and the desired states of the chosen emotions. The speaker was then left for a period of time to get familiar with the script and the recording software tool and to prepare their emotional acting. In the next stage, the speaker was asked to make a short recording test to adapt to the recording equipment. No time limitation was imposed on the preparation session. The speakers could start the recording session when they were fully prepared.

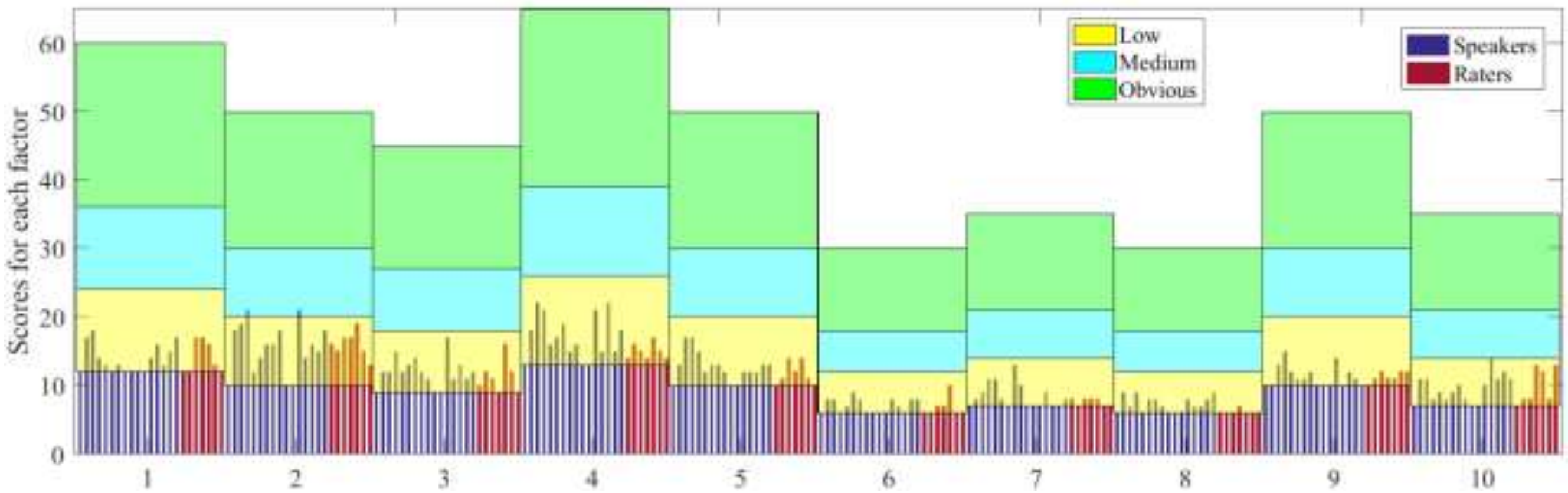

Fig. 8. Scores for the speakers and raters in each of the factors in SCL-90.

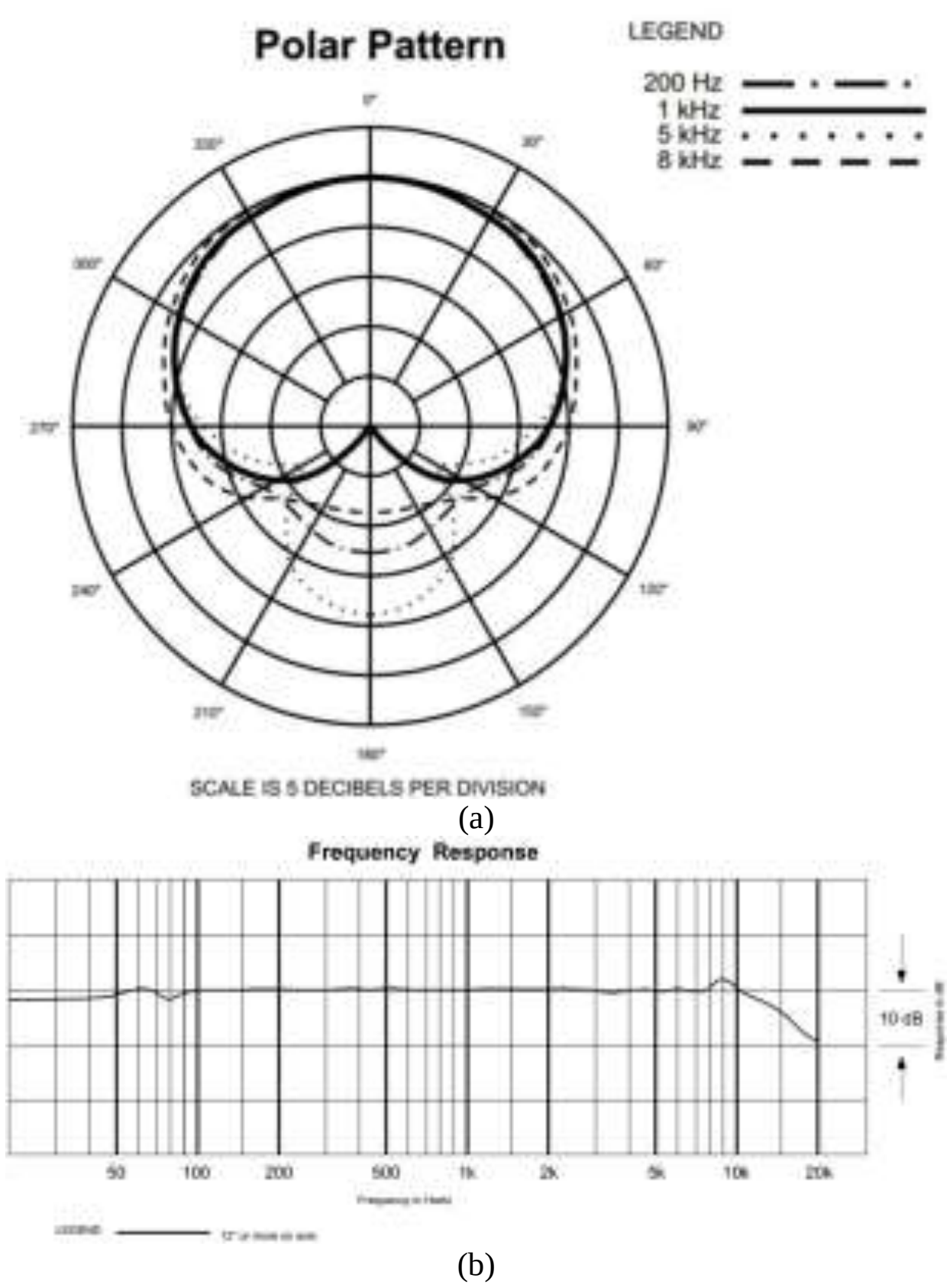

Fig. 9. Characteristics of the AT2020USB microphone: (a) Polar Pattern (b) Frequency Response

In the recording session, each speaker set a personal code with 3 letters before recording. The rest of the recording process was guided by a graphic interface that presented the assigned sentences and the desired emotion (in Chinese), as shown in Fig. 10. Each speech sample was played out immediately after recording. After judging the voice quality and validity of intended emotion, the speaker could decide whether to progress

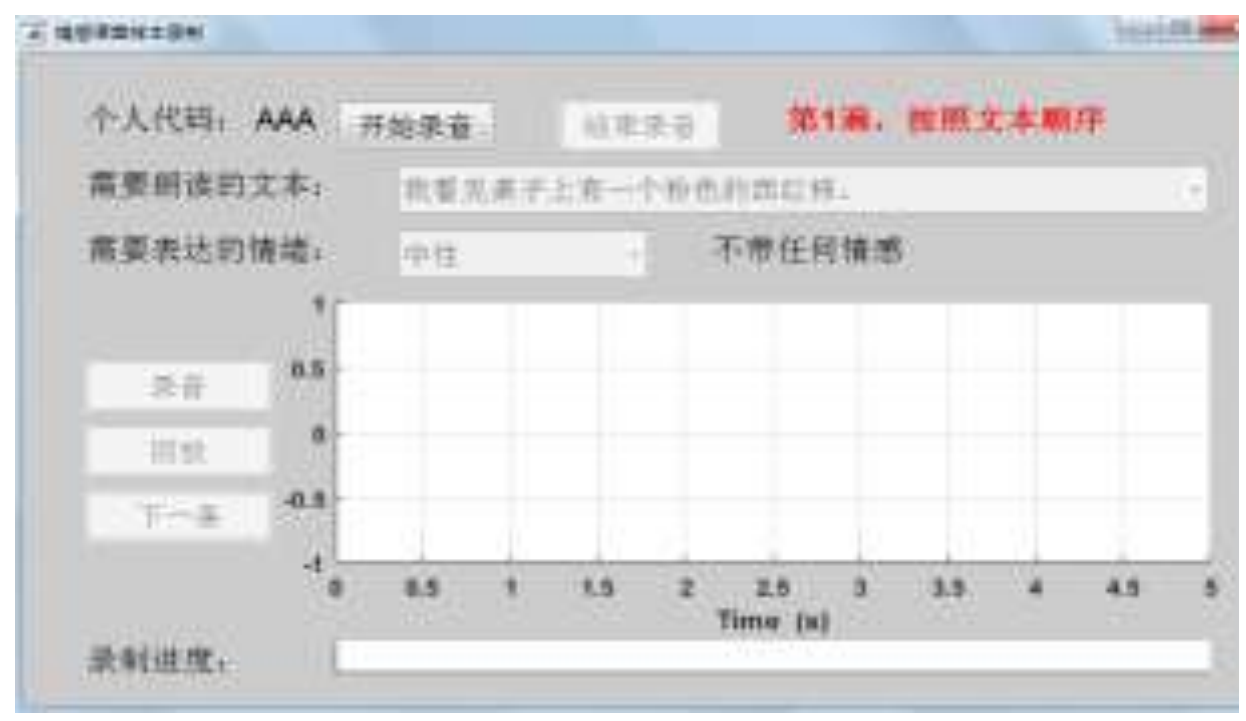

Fig. 10. Graphic interface in recording session.

to the next item or to record the current item again.

The recording session was divided into 3 rounds. In the first round, the speakers read each sentence in the script in all 7 emotional states before proceeding with the next sentence. This first round thus allows the speakers to record different emotional states using the same verbal content. In the second round, the speakers were requested to maintain one emotional state over all 16 sentences before passing to another emotional state. This round thus allows the speakers to maintain a relatively stable emotion over a period of time. In the third round, the sentences and emotions appeared in a random sequence to avoid the influence of fatigue due to repeated sentences or emotions. As a result, for each speaker, 3 speech samples were collected for each emotional state for each sentence. This results in 16 (speakers) × 16 (sentences) × 7 (emotions) × 3 (times) = 5376 emotional speech samples.

All the speech samples were recorded monaurally with a sampling frequency of 44.1kHz at 16-bit precision. The file names of the recorded speech samples were set automatically according to the following rules. The first letter refers to the gender of the speaker, the next three letters are the unique speaker code, then the letter “t” preceding the sentence index, a letter “e” preceding the emotion index, and the letter “v”

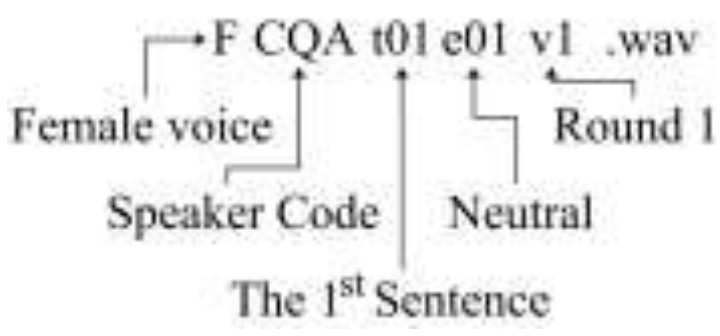

Fig. 11. Example of file name setting.

preceding the recording round index. An example of the file naming is shown in Fig. 11.

## E. Emotion rating to recorded speech samples

The evaluation of the speech samples was performed by several raters independently. All raters were also university students who have the same living and communication environment as the speakers in the recording process. They were selected also after the SCL-90 test as explained in the previous section. None of the raters participated in the recordings. Therefore, they evaluated the speech samples only as observers.

The raters were asked to mark the perceived emotions for the speech samples on a VA coordinate after listening to the speech samples using a software GUI, resulting in one coordinate for each speech sample. They were allowed to listen to the speech samples as many times as necessary to form their confident judgments. The position of neutral speech is defined at the origin in the VA coordinate system. The numerical ranges in both valence axis and arousal axis are defined as [-3, 3], where, for the raters' reference, the integer values can be seen as boundaries of emotion intensity.

In some existing emotional speech datasets, the rating results of emotions were also used as a screening of the speech samples; samples on which the raters disagreed concerning the emotional states were deleted to ensure the typicality and consistency of the labeled emotions. In our dataset, moderate emotions were collected in addition to intense ones; thus, the typicality of emotions is actually no longer an essential factor as in some of the existing datasets. Emotion is a subjective state expressed and perceived by people. Individual differences express different levels of emotion intelligence. Through variations of expression modes, speakers can display different levels of emotion expressivity, whereas raters as listeners can also exhibit unequal skill of emotion understanding with respect to intended emotions expressed by speakers. Finally, the divergence between emotion expression and perception characterizes the emotion discrepancy which commonly exists in real life and is a major reason for misunderstanding between people.

Given the fact that the speakers in the recording process were allowed to repeat the same item as many times as desired when listening to the playback speech, we can assume that the saved speech samples already faithfully reflected the emotions intended by the speakers and thereby no speech samples should be deleted from the dataset. The rating results by the raters are used to reflect listeners' emotional perceptions.

A key question in human emotion rating is the number of raters required in order to ensure the quality and diversity of the ratings. Coding results from too few raters might present severe individual bias and do not reflect the diversity of rating results, while using too many raters might lead to enormous workload. In this work, we studied this issue and concluded that using seven raters is enough to reach a reliable rating of the emotional speech samples.

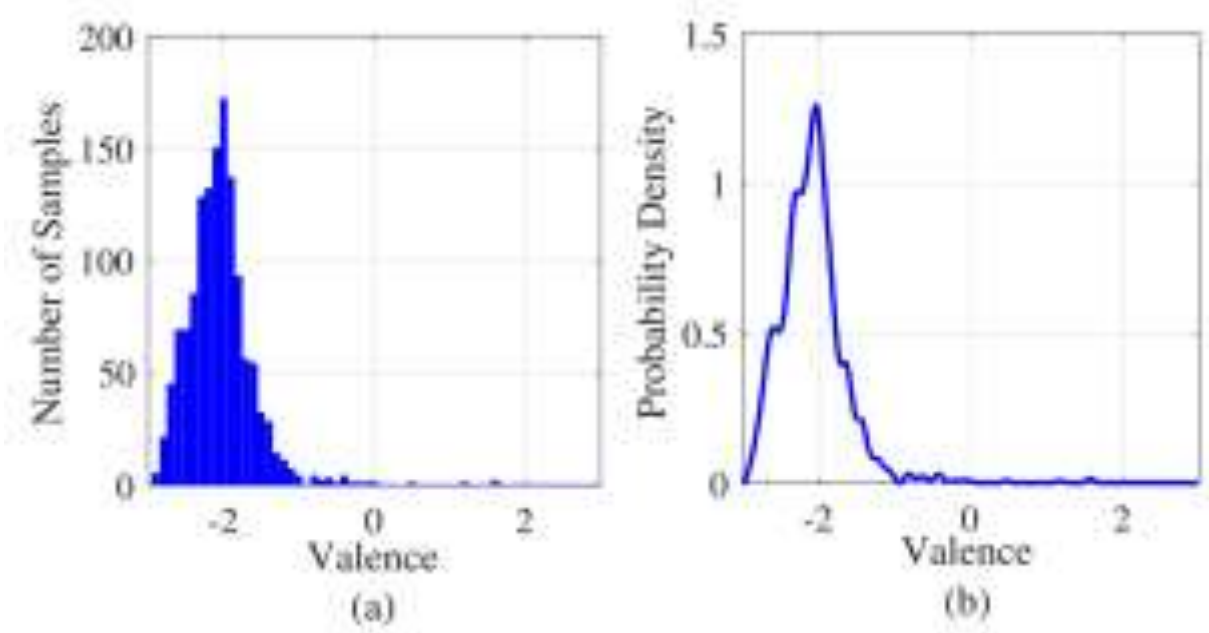

Fig. 12. Histogram and distribution of rated coordinate values (a) Histogram (b) Probability Density

Indeed, according to central limit theorem, when the number of raters $n$ is large enough, the distribution of average rating results from these raters becomes increasingly a normal distribution. We can thus compare the actual distribution of the average rating results with the theoretical one and set the number of raters $n$ to minimize the discrepancy between the actual and theoretical distributions.

Specifically, each rater is arbitrarily assigned an index number. The distribution of the mean rating for a given emotional state from $k$ raters is evaluated using the ratings from raters No. 1 to No. $k$. For the $x^{th}$ emotional state, we calculate from $k$ raters the mean value $\mu_{Dxk}$ and the variance$\sigma_{Dxk}$, with D referring to the one of the two axes of the dimensional emotion space, *i.e.*, A for Arousal and V for Valence, $\mu_{Axk}$ or $\mu_{Vxk}$, and then instantiate the corresponding theoretical normal distribution denoted as the the base probability density $Bpd_{xk}(D)$ for both arousal dimension $Bpd_{xk}(A)$ and valence dimension $Bpd_{xk}(V)$:

$$Bpd_{xk}(D) = \frac{1}{\sqrt{2\pi}\sigma_{Dxk}} e^{-\frac{(D-\mu_{Dxk})^2}{2\sigma_{Dxk}^2}} \quad (1)$$

Where the parameter D refers to Arousal or Valence dimension.

The actual probability density for a given emotional state from the $k$ raters is estimated by means of histograms. The rating range, from -3 to 3, is divided into intervals with 0.1 width, and the number of samples falling in each interval is then computed, resulting in a histogram as illustrated in Fig. 12 (a). Then the histograms for all the emotional states are smoothed and normalized into a probability density, as illustrated in Fig. 12 (b). They are used as actual probability density $Apd_{xk}(D)$ in contrast with the theoretical one as expressed in Eq.(1), with $D$ referring to one of the two axis of the dimensional emotion space, *i.e.*, arousal or valence, the subscript $x$ to the $x^{th}$ emotional state, and $k$ to the rating results from the $k$ raters.

Fig. 13 illustrates the theoretical base probability densities and actual probability densities of mean ratings of 2, 3 and 7

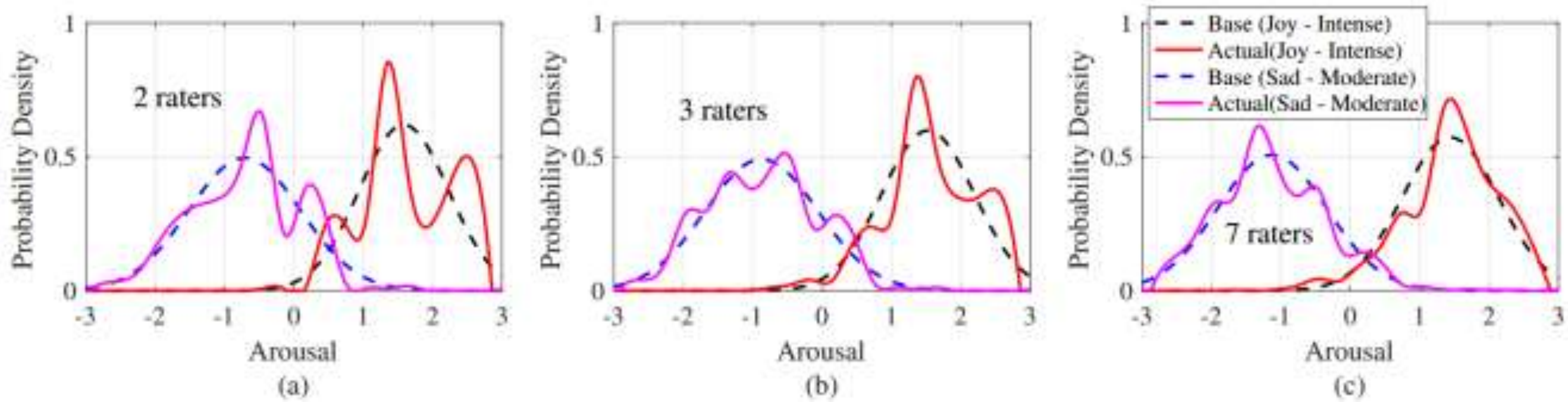

Fig. 13. Probability densities of proximal labels on moderate anger and intense sad by different number of raters in Arousal dimension (a) 2 raters (b) 3 raters (c) 7 raters. Black dashed line and red solid line for base and real probability densities of intense joy, blue dashed line and magenta solid line for base and real probability densities of moderate sad.

raters, respectively, for the two emotional states, *e.g.*, intense joy and moderate sad in the Arousal axis. As can be seen in Fig. 13, with the increase of number of raters, the actual probability densities get smoother, and closer to the theoretical base probability densities. We can calculate the Mean Square Root Errors (MSREs) between the base probability densities and actual ones of $k$ raters for the $x^{th}$ emotional state $ME_{Dxk}$, *i.e.*, $ME_{Axk}$ for Arousal and $ME_{Vxk}$ for Valence, to evaluate the degree of approximation between the two distributions:

$$ME_{Dxk} = \sqrt{\frac{1}{6}\int_{-3}^{3} |Bpd_{xk}(D) - Apd_{xk}(D)|^2 dD} \qquad (2)$$

In order to have all MSREs within a same range of values for all the emotional states and thereby facilitate their interpretation, all MSREs are normalized with respect to the MSRE from only one rater, and denoted as $NE_{Dxk}$ for the $x^{th}$ emotional states from $k$ raters ($NE_{Axk}$ for Arousal and $NE_{Vxk}$ for Valence):

$$\begin{cases} NE_{Dx1} = 1 \\ NE_{Dxk} = \dfrac{ME_{Dxk}}{ME_{Dx1}}, \quad k > 1 \end{cases} \qquad (3)$$

The difference $DE_{Dxk}$ in normalized MSRE for the $x^{th}$ emotional state between $k-1$ raters and $k$ raters can be calculated using Eq.(4), with $DE_{Axk}$ for Arousal and $DE_{Vxk}$ for Valence:

$$DE_{Dxk} = NE_{Dx(k-1)} - NE_{Dxk} \qquad (4)$$

Fig. 14 plots the normalized MSREs (Fig. 14.(a) and (c)) and their differences (Fig. 14.(b) and (d)) for different emotional states between the theoretical base distributions and the actual ones. As can be seen in Fig. 14, (a) for Arousal and (c) for Valence, with the increase of number of raters, the overall trends in the normalized MRSEs are decreasing, with occasional increases when the number of raters are less than 4. When the number of raters reaches 6 or 7, the normalized mean square root errors converge to relatively stable values.

As for the differences of normalized MSREs plotted in Fig.14.(b) for arousal and (d) for valence and listed in Table. IX, they reach to a range very close to zero for all the emotional states in both dimensions when the number of raters is between 6 and 7.

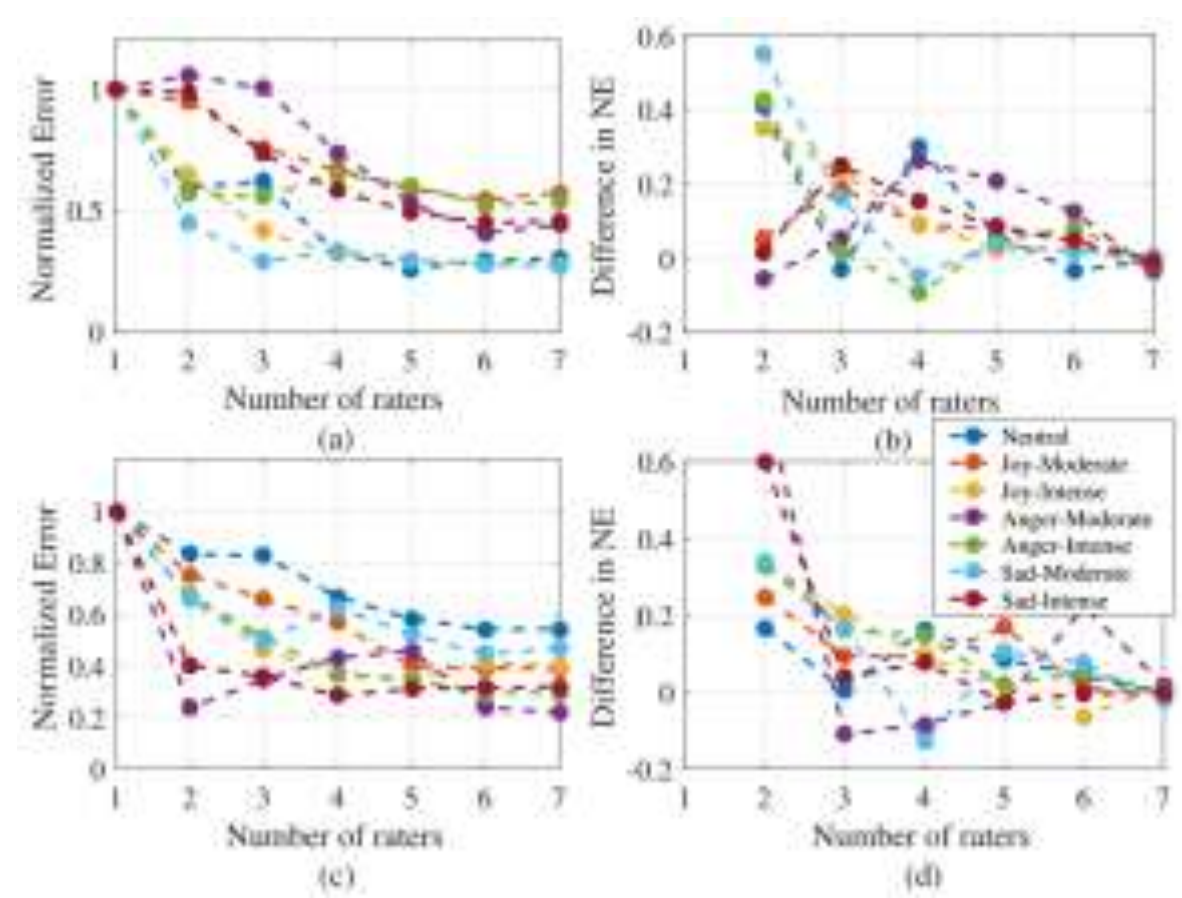

Fig. 14. Normalized mean square root Errors (MSREs) between the base distributions and real distributions, with different number of raters (a) Normalized Error in Arousal dimension (b) The difference in normalized error, in Arousal dimension between rent number of raters (c) Normalized Error in Valence dimension (d) The difference in normalized error, in Valence dimension

With the differences of normalized MSREs close to zero, this means the actual probability densities and base probability densities could not get closer when the $7^{th}$ rater is added to the subjective rating. Thus, we assume 7 independent raters are enough to give out reliable subjective rating results on emotions as proximal labels for our dataset.

## IV. Analysis of Rating Distribution and the Ratings' Consistency

A key question in this subjective annotation of emotional speeches is the consistency of the ratings from different raters. In this section, we analyze the major features of the rating distributions to gain insights into the overall trends of the ratings.

### *A. Overall distribution of emotional rating*

The emotional speeches have been annotated by seven raters in the dimensional valence/arousal space. Fig. 15 depicts the overall distribution of their ratings. Each emotional state is represented by an oval in Fig. 15, where the centers correspond to the mean value of the ratings, and the radii in the two dimensions correspond to the standard deviation in the valence and arousal axis.

TABLE IX
DIFFERENCE IN NORMALIZED MEAN SQUARE ERRORS BETWEEN 6 RATERS AND 7 RATERS

| Emotional State | Neutral | Moderate Joy | Intense Joy | Moderate Anger | Intense Anger | Moderate Sad | Intense Sad |
|---|---|---|---|---|---|---|---|
| Arousal | -0.0042 | -0.0307 | -0.0056 | -0.0338 | -0.0146 | 0.0076 | -0.0087 |
| Valence | -0.0001 | 0.0034 | 0.0146 | 0.0182 | 0.0007 | -0.0221 | -0.0038 |

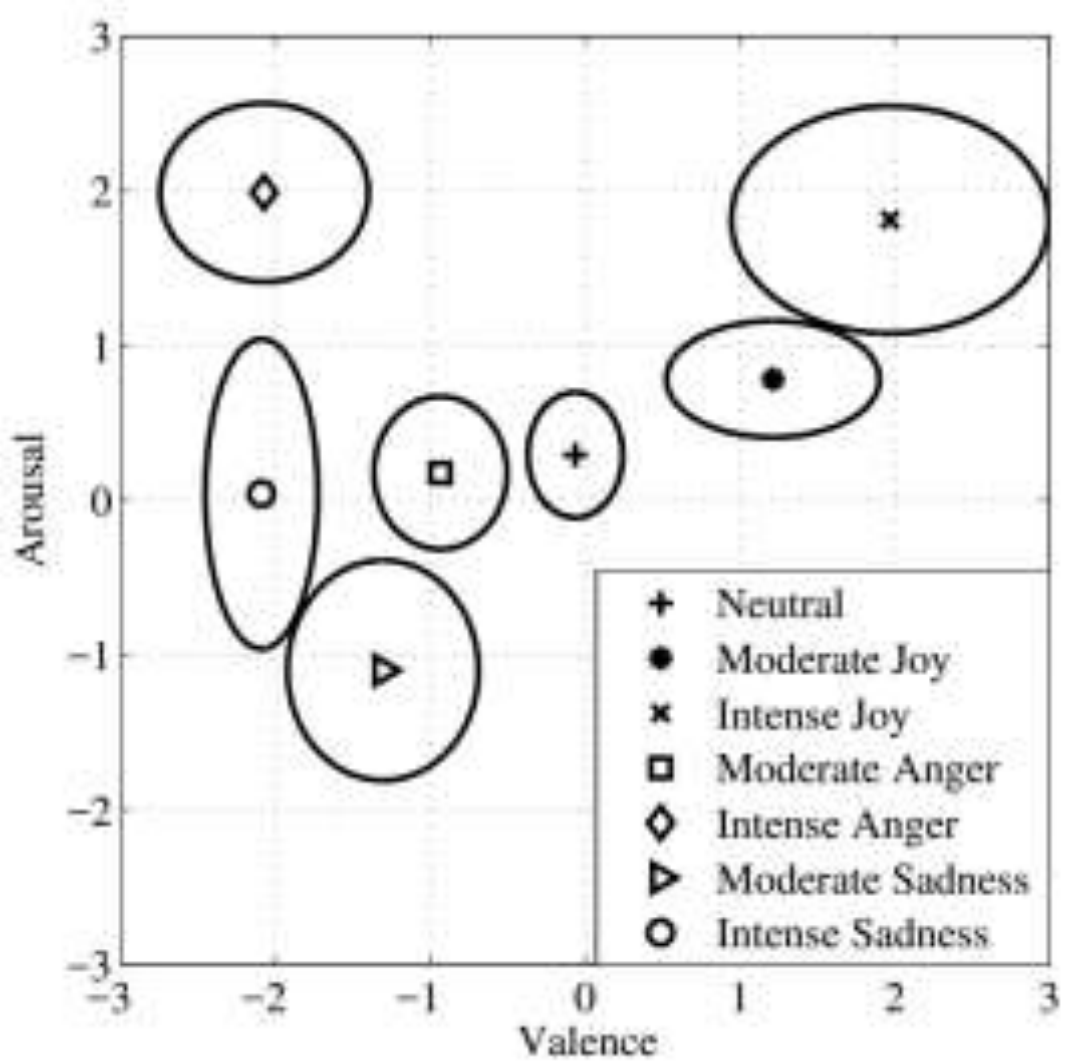

Fig. 15. Overall distribution of emotional ratings. Different Markers stand for different emotional states.

The first observation over the rating distributions is that the neutral emotion is not exactly located at the origin of the VA coordinate system but near it with relatively small variances in both dimensions.

Intense anger is almost symmetric to intense joy in the valence dimension, with high positive arousal and high negative valence, while moderate anger presents a very small positive value in arousal and a low negative value in valence. Because the center position of neutral emotion also has a small positive value in arousal, we can see that moderate anger can be only distinguished from neutral speech in the valence axis. Note that the areas for intense anger and moderate anger are quite far away to each other, which indicates that the expression way in anger could change significantly with the increase of intensity. On the other hand, the ratings of moderate anger speech samples are very close to neutral and moderate sadness, which could be a source of confusion.

Moderate sadness is almost symmetric to moderate joy with respect to the origin of the VA coordinate, with negative values in both dimensions. The distribution of intense sadness is more negative in valence dimension, while in the arousal dimension, the center is also very close to zero as in moderate anger and is distributed over a relatively large range in the arousal dimension, from negative to positive.

### *B. Consistency between raters' perception and speakers' intention*

One of the key questions in evaluation of the quality of a speech dataset is the reliability of the subjective ratings or how the emotions perceived by the raters and the ones intended by the speakers are consistent each other. Zhao et al. [52] analyzed 22 state-of-the-art reliability indexes and concluded that all existing indexes suffer from some unrealistic assumptions and paradoxes. Different reliability indexes display different bias according to a number of factors, including the number of class categories and/or balance of class samples. As a result, the index to evaluate the reliability of raters should be chosen according to specific conditions of a given application to avoid significant bias. In our work, the subjective rating involves 7 emotional categories, with even distribution of distal labels in each category, the Cohen's Kappa coefficient ($\kappa$) [53] [54] (or its equivalent Rogot & Goldberg's A2 [55]) is recommended for use. In denoting the distal emotional labels (speakers' intention) as case A, and the judgment according to a rater as case B, the Cohen's $\kappa$ is defined in Eq.(5):

$$\kappa = \frac{P_o - P_e}{1 - P_e} \tag{5}$$

where $P_e$ refers to the hypothetical probability of chance agreement, or prior probability. This prior probability on the 7 emotional categories can be calculated by Eq.(6) :

$$P_e = \sum_{k=1}^{7} P_{Ak} P_{Bk} \tag{6}$$

where $P_{Ak}$ refers to the probability of speakers assigning speech samples as the $k^{th}$ emotional state. In our case, each of the 7 emotional state has the same number of samples, we have thus $P_{Ak} = 1/7$ for any $k$. Similarly, $P_{Bk}$ refers to the prior probability of raters assigning a speech sample as the $k^{th}$ emotional state and equals to 1/7.

$P_o$ refers to the observed agreement, or posterior probability and can be calculated on the 7 emotional categories by Eq.(7):

$$P_o = \sum_{k=1}^{7} \frac{N_{kk}}{N} \tag{7}$$

where $N_{kk}$ is the number of speech samples assigned as the $k^{th}$ emotional state by both speakers and raters, and $N$ is the total number of all speech samples, *i.e.*, 5376 in MES-P.

The posterior probability generally is assumed to be no smaller than a chance level, Cohen's $\kappa$ lies then in the range [0,1]. $\kappa$ = 0 corresponds to the situation when $P_o$ equals the chance level Pe, suggesting that ratings are not consistent or reliable at all. On the contrary, $\kappa$ = 1 corresponds to the situation where $P_o$ equals 1, indicating that ratings exhibit perfect consistency or they are completely reliable. In dividing the value range of Cohen's Kappa coefficient into 5 areas, different consistency descriptions can be assigned as shown in Table X in following [56].

TABLE X
DESCRIPTION OF CONSISTENCY ACCORDING TO COHEN'S KAPPA COEFFICIENT

| Kappa | 0~0.2 | 0.2~0.4 | 0.4~0.6 | 0.6~0.8 | 0.8~1 |
|---|---|---|---|---|---|
| Consistency | Slight | Fair | Moderate | Substantial | Almost perfect |

In MES-P, the emotions expressed by the speakers are based on discrete emotional states during the recording process whereas raters' perceived emotions are evaluated based on the continuous coordinate values in the dimensional VA space, a conversion criterion is needed to define the emotional states from the continuous emotion rating values in order to be able to assess the consistency of labels between speakers and raters.

The criterion for inferring discrete emotional states from the continuous proximal labels is based on the nearest neighbor classifier in the V-A space. Specifically, the center positions of the 7 emotional states are set as reference points for each emotional state. Given a labeled speech sample, one can thus compute the Euclidean distances to these 7 centers, as in Eq. (8):

$$D_{ijkm} = \sqrt{\left(V_{ijk} - C_{V_m}\right)^2 + \left(A_{ijk} - C_{A_m}\right)^2}, m = 1\sim 7 \quad (8)$$

Where $V_{ijk}$ and $A_{ijk}$ are the coordinates for the $k^{th}$ speech sample in the $j^{th}$ emotional state (according to the speakers' intention) by the $i^{th}$ rater, whereas $C_{V_m}$ and $C_{A_m}$ refer to the centers for the $m^{th}$ emotional state in Valence and Arousal dimensions, respectively. Thus, 7 distances for each speech sample are obtained.

Each speech sample with a given proximal label is then classified as the emotional state m whose center position $C_{V_m}$ and $C_{A_m}$ is the closest. As a result, each rating expressing the raters' emotion perception can be assigned a discrete emotional state.

The Cohen's Kappa coefficients between distal labels and the proximal ones, *i.e.*, subjective judgments of each of the 7 raters are listed in Table XI. Among the Cohen's $\kappa$ from 7 raters, 5 values are above 0.8, which suggests almost perfect consistency, and the other 2 are over 0.75, which means substantial consistency, according to Table X. The average Kappa coefficient over the 7 raters is around 0.84, thereby exhibiting almost perfect consistency.

The Cohen's Kappa coefficients between the raters' perceptions and the speakers' intentions were also calculated for each of the 7 emotional states, as shown in Fig. 16. As can be seen in Fig. 16, neutral is the easiest emotional state where all the raters exhibit Kappa coefficients as high as 0.9. It is then followed by intense and modest anger (emotion index 4 & 5) for which all the raters display a Kappa coefficient above 0.8, thus suggesting almost perfect consistency. For the emotional family joy (emotion index 2 & 3), 2 raters gave subjective

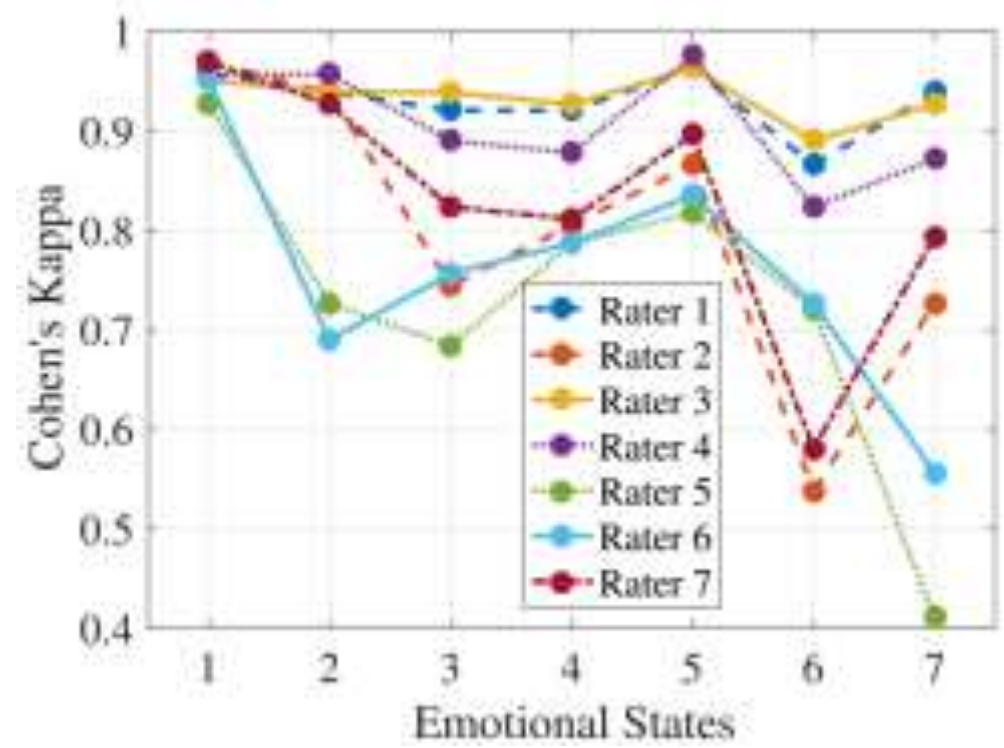

Fig. 16. Cohen's Kappa coefficients on different emotional state. Meaning of emotional index: 1-Neutral, 2-Moderate joy, 3-Intense joy, 4-Moderate anger, 5-Intense anger, 6-Moderate sadness, 7-Intense sadness

ratings with Kappa coefficients around 0.7, which indicate weaker consistence than neutral and anger. The sad family shows the worst consistency; only 3 raters exhibit Kappa coefficients above 0.8, and there are even two raters display Kappa coefficients smaller than 0.6 for moderate sad (emotion index 6), and two other raters for intense sad (emotion index 7). These low Kappa coefficients suggest that these emotional states are quite hard to be recognized by the raters.

### C. *Consistency between raters*

The previous subsection shows that, depending upon the underlying emotional state, different raters depict different Kappa coefficients, thereby suggesting that raters provide proximal labels differently. In this subsection, we further investigate this issue and highlight the consistency between raters through the Cohen's Kappa coefficient. For this purpose, case A and B defined in Eq. (5) through (7) are now substituted by ratings from 2 different raters, respectively. The probability of chance agreement $P_e$ remains 1/7, and $P_o$ is defined as the posterior probability that a pair of raters' rating results agrees in terms of emotional state on speech samples.

Table XII synthesizes the results of this analysis. As can be seen in Table XII, the consistency between raters varies a lot with the Kappa coefficient, ranging from 0.6649 to 0.9375, suggesting that the consistency of ratings between pairs of raters varies from substantial to almost perfect. It is noticeable that rater 5 and 6, who both exhibit relatively lower emotional consistency with respect to speakers' intentions, display a $\kappa$ as high as 0.8542 to each other. This result suggests that these 2 raters have similar manner of emotion perception, while exhibiting large difference of their emotion perception with respect to speakers' emotion intentions.

## V. ANALYSIS OF HUMAN EMOTION EXPRESSION ABILITY AND PERCEPTION SKILL: DISTAL VS. PROXIMAL

The analysis in the previous section shows that there is highly emotional consistency between speakers and raters, and between different raters. However, we also discover that vocal emotion is expressed and perceived differently by different people who exhibit different level of emotion intelligence. In this section, using the two sets of emotional labels in MES-P, namely distal and proximal labels, we evaluate the speakers' ability of expressing vocal emotions and raters' emotional perception sensitivity.

TABLE XI
COHEN'S KAPPA COEFFICIENTS BETWEEN EACH RATER AND SPEAKERS

| Rater | 1 | 2 | 3 | 4 | 5 | 6 | 7 | Average |
|---|---|---|---|---|---|---|---|---|
| Cohen's Kappa | 0.9054 | 0.8299 | 0.9332 | 0.9157 | 0.7578 | 0.7509 | 0.8064 | 0.8430 |

TABLE XII
COHEN'S KAPPA COEFFICIENTS BETWEEN EACH RATER AND SPEAKERS

| Rater | 1 | 2 | 3 | 4 | 5 | 6 | 7 |
|---|---|---|---|---|---|---|---|
| 1 | 1 | 0.7839 | **0.9375** | 0.8724 | 0.7161 | 0.7309 | 0.7674 |
| 2 | | 1 | 0.7943 | 0.8281 | 0.6658 | 0.6780 | 0.8984 |
| 3 | | | 1 | 0.8898 | 0.7257 | 0.7335 | 0.7778 |
| 4 | | | | 1 | 0.7422 | 0.7439 | 0.8073 |
| 5 | | | | | 1 | 0.8542 | 0.6649 |
| 6 | | | | | | 1 | 0.6701 |
| 7 | | | | | | | 1 |

TABLE XIII
ACCURACY IN EXPRESSION BY EACH SPEAKER (%)

| Speaker | 1 | 2 | 3 | 4 | 5 | 6 |
|---|---|---|---|---|---|---|
| Accuracy | 65.60 | 89.58 | 83.83 | 93.96 | 92.79 | 91.26 |
| Speaker | 7 | 8 | 9 | 10 | 11 | 12 |
| Accuracy | 93.77 | 90.75 | 65.40 | 84.61 | 87.82 | 91.93 |
| Speaker | 13 | 14 | 15 | 16 | Average | Std |
| Accuracy | 82.95 | 90.84 | 90.26 | 89.58 | 86.54 | 8.57 |

TABLE XIV
CONFUSION MATRIX IN SPEAKERS' EXPRESSION (%)

| | A-I | A-M | J-I | J-M | N | S-I | S-M |
|---|---|---|---|---|---|---|---|
| A-I | **93.38** | 2.19 | 0.04 | 3.47 | 0.04 | 0.78 | 0.09 |
| A-M | 5.36 | **90.91** | 1.00 | 2.43 | 0.17 | 0.09 | 0.04 |
| J-I | 1.48 | 7.77 | **86.26** | 2.02 | 2.43 | 0.00 | 0.04 |
| J-M | 8.12 | 0.3 | 1.43 | **82.29** | 0.39 | 4.69 | 2.78 |
| N | 0.82 | 0.43 | 0.26 | 1.48 | **92.74** | 1.66 | 2.60 |
| S-I | 1.69 | 0.04 | 0.04 | 5.86 | 0.13 | **85.62** | 6.61 |
| S-M | 0.22 | 0.00 | 0.00 | 3.69 | 13.11 | 8.16 | **74.83** |

### *A. Emotion expression ability of speakers*

Different people exhibits different facility for expressing vocal emotions. People with better emotion expression ability will make easily understandable by other people the vocal emotion he/she wants to express. With the proximal labels, we propose to evaluate emotion expression ability of speakers in terms of average emotion recognition accuracy of all the raters.

Specifically, given a speaker and his or her emotional speech samples, we compute for each rater his emotion recognition rate using the proximal labels and then average the emotion recognition rate over all the 7 raters. Table. XIII synthesizes the average emotion recognition rate for each speaker along with the average recognition accuracy over all speakers and its standard deviation. As can be seen in Table. XIII, the 16 speakers exhibit different emotion expression ability. With an average recognition accuracy of 93.96% and 93.77% by the raters, speaker 4 and 7 display the best emotion expression ability. The worst ones in terms of emotion expression ability are speaker 1 and 9 who only display an average recognition rate of 65.60% and 65.40%, respectively. The mean recognition accuracy over all the speakers is 86.54% with a standard deviation of 8.57, which shows obvious difference in terms of emotion expressing ability among the speakers. 10 speakers out of 16 get over 90% of their speech samples correctly recognized by the raters.

The confusion matrix of speakers' emotion expression ability (average over 16 speakers) is listed in Table. XIV. The best expressed state is intense anger (A-I), with a recognition rate of 93.38% by all raters, while the worst expressed emotional state is moderate sadness (S-M), which displays a recognition rate of only 74.83%. It is much confused with neutral for 13.11%. A noticeable confusion within the sadness emotional family also appears. Surprising confusion patterns occur on two pairs of emotional states. First, high confusion occurs between moderate joy (J-M) and intense anger (A-I), around 5%. Second, both moderate anger (A-M) and moderate joy (J-M) tend to be misjudged as intense (A-I) anger, with a confusion rate of 5.36% and 8.12%, respectively.

### *B. Emotion perception sensitivity of raters*

Symmetric to a speaker's emotion expression ability is a listener's emotion perception sensitivity. A person with high emotion perception sensitivity is supposed to have the ability to recognize with high accuracy vocal emotions across speakers and even though the expressed emotions were moderate ones. Thanks to the distal and proximal labels, we can quantify raters' emotion perception ability using the recognition accuracy of each rater over all emotional speech samples in MES-P.

Specifically, we consider as the ground truth the distal labels, *i.e.*, intended emotions conveyed by the emotional speech samples from speakers, and proximal labels, *i.e.*, ratings of each rater as the classification results of a human classifier and derive an emotion recognition rate for each rater. Table. XV lists the emotion recognition accuracy for each rater. As can be seen in Table. XV, different raters exhibit variable emotion sensitivity. With an accuracy of 94.27%, rater 3 displays the best emotion sensitivity whereas rater 6, with an accuracy of 78.65%, exhibits the worst one. The average emotion recognition over all the raters is 86.54%, with a standard deviation of 6.07%.

TABLE XV
ACCURACY BY HUMAN RATERS (%)

| Rater | 1 | 2 | 3 | 4 | 5 | 6 | 7 |
|---|---|---|---|---|---|---|---|
| Accuracy | 91.89 | 85.42 | 94.27 | 92.93 | 79.24 | 78.65 | 83.41 |
| Average | 86.54 | Std | 6.07 | | | | |

TABLE XVI
AVERAGE CONFUSION MATRIX BY HUMAN RATERS (%). LETTERS A, J, S, N REFER TO EMOTIONS AS ANGER, JOY , SADNESS, NEUTRAL, AND LETTERS I AND M REFER TO INTENSE AND MODERATE EMOTIONS RESPECTIVELY.

| | A-I | A-M | J-I | J-M | N | S-I | S-M |
|---|---|---|---|---|---|---|---|
| A-I | 91.74 | 4.32 | 0.07 | 0.30 | 1.64 | 1.71 | 0.22 |
| A-M | 0.30 | 86.68 | 0.00 | 0.00 | 8.48 | 1.26 | 3.27 |
| J-I | 0.67 | 0.82 | 84.6 | 11.09 | 2.83 | 0.00 | 0.00 |
| J-M | 0.00 | 1.12 | 4.76 | 87.35 | 6.32 | 0.15 | 0.30 |
| N | 0.00 | 2.90 | 0.07 | 0.30 | 95.83 | 0.07 | 0.82 |
| S-I | 5.43 | 3.05 | 0.00 | 0.00 | 0.37 | 83.85 | 7.29 |
| S-M | 0.30 | 6.85 | 0.00 | 0.00 | 3.65 | 13.47 | 75.74 |

Table. XVI is the confusion matrix by human raters where the accuracy for each emotional state is averaged over all the 7 raters. As can be seen in Table. XVI, the most noticeable confusions occur within emotional families, *e.g.*, between moderate and intense joy (J-M vs. J-I), or between moderate and intense sadness (S-M vs. S-I). Furthermore, moderate anger (A-M) is also much confused with neutral (N). Looking back at Fig.15, this confusion is quite understandable as ratings for moderate anger and neutral are the closest emotional state in the VA space.

## VI. BASELINE PERFORMANCE ON MES-P FOR EMOTION RECOGNITION

In this section, we provide a baseline performance for automatic emotion recognition based on the proposed MES-P dataset. Because MES-P has both distal and proximal labels, the training of a predictive model can be carried out on distal labels but tested on proximal ones or inversely. Section VI-A defines the experimental settings. Section VI-B introduces the feature set. Section VI-C discusses the results.

### A. Experimental settings

Given the 7 emotional states from 3 emotional families and neutral speech, we define 2 classification problems, subsequently referred to as *A* and *B*, respectively. Problem *A* defines a setting similar to those in previous work and enables comparisons with the state of the art. Specifically, problem *A* only considers the intense emotions with neutral as most previous datasets do and results in a classification problem of 4 classes. Problem *B* considers all the emotional states and results in a classification problem of 7 classes.

TABLE XVII
CLASSIFIER LIST AS MODEL CODES

| Training | Evaluation | Model Code |
|---|---|---|
| Distal | Distal | D-D |
| | Proximal | D-P |
| Proximal | Distal | P-D |
| | Proximal | P-P |

Thanks to distal and proximal labels, possible perception distortions between intended or perceived emotions upon automatic classifications can be investigated now. Specifically, for each classification problem, *A* or *B*, we build two different classifiers according to the training data labels being distal or proximal. They can be further evaluated in two ways, using distal or proximal labels. Table XVII summarizes these variants with each model encoded by two letters as "*T-E*", where *T* refers to the label type (D for Distal, and P for Proximal) used in training, whereas *E* refers to the label type used in the evaluation. For example, model code D-P refers to a predictive model trained with distal labels, but its predictions evaluated with proximal labels.

In analogy to the evaluation of human emotion intelligence in Section V, model D-D enables to quantify how similarly a predictive model trained on distal labels encodes its emotions as speakers, whereas model D-P corresponds to the emotion expression ability of the learned predictive model. Conversely, model P-D expresses the emotion sensitivity of a predictive model trained on proximal labels whereas model P-P enables to quantify how similarly a predictive model trained with proximal labels decodes its emotions as human raters.

However, each emotional speech sample is rated by 7 raters and results in 7 VA coordinates in the VA space. We need to assign a unique discrete emotion label for each of these emotional speech samples. For this purpose, we first compute the average value of the 7 VA coordinates and convert this continuous average rating into a discrete proximal label based on the nearest neighbor as in Section IV-B using Eq. (8).

Given the limited number of speakers, *i.e.*, 16, classifiers are evaluated with a leave-one-speaker-out manner. For the emotion prediction of speech samples from the leaving out speaker, the classifiers are trained with the speech samples from all the other speakers. The classification accuracy and the confusion matrix are then averaged over all the models with each speaker used once as leave-out speaker.

The training of predictive models are based on SVM using the WEKA platform [57], with the SMO (Sequential Minimal Optimization) function. The default SMO parameters in WEKA are adopted, with the default polynomial kernel.

### B. Feature set

In this work, we adopt the baseline feature set of the INTERSPEECH 2013 computational paralinguistic challenge [58], which features 4 sub-challenges, namely, social signals, conflict, emotion, autism. This feature set is also the standard feature set for the INTERSPEECH 2017 computational paralinguistic challenge on addressee, cold, and snoring [59]. The features in this set are extracted form 64 low-level descriptors (LLD) and their delta coefficients, with a series of functions applied to these LLDs and $\Delta$LLDs, resulting into 6372 static features. The LLDs and the functions applied to them are listed in Table XVIII. The features are extracted using TUM's open-source openSMILE feature extractor [60], with the configuration IS13_ComParE.conf. The features are normalized before feeding into learning models. No feature selection was performed in this work.

### *C. The results*

The overall experimental results are synthesized in Table XIX. The average human perception accuracies for the 2 prob-lems are also reported in the last column. In the following, we discuss in detail the experimental results of the 2 classification problems.

*1) Problem A - 4-class classification on intense emotions and neutral:* In this classification problem, only neutral speech samples and those with intense emotions are used for training and testing. This setting is expected as the easiest one given the pronounced nature of emotions. It is comparable to the emotions considered in existing early datasets, *e.g.*, Emo-DB[24], or DES [21].

The corresponding classification accuracies on MES-P are listed in the first row of Table XIX. As can be seen in Table XIX, 2 predictive models have been trained using either distal (D) or proximal (P) labels. They are further evaluated using these two types of labels and result in 4 classification accuracies, namely D-D, D-P, P-D, P-P, which differ only slightly, from 66.60% to 67.58%. These results suggest that the trained predictive models depict similar encoding (D-D accuracy) and decoding skill (P-P accuracy). Furthermore, they also exhibit similar emotion expression ability (D-P accuracy) and emotion sensitivity (P-D accuracy).

For comparison, using the same SVM with the default parameters on Weka as well as the same leave-one-speaker-out protocol, we also perform training and classification of similar emotional states on DES [21], which is a widely used public emotional speech dataset for years, and achieve 50.12% accuracy. This result comforts us in the emotion expression quality of the proposed MES-P dataset.

The confusion matrices are given in Fig. 17. As can be seen in Fig. 17, with X being letter D or P, the predictive model D-X trained using distal labels predicts better sadness than the one P-X trained using proximal labels; conversely, the predictive model P-X better predicts anger than the model D-X. For all the cases, the highest confusions occur between anger and joy, which are consistent with the early studies in the literature [61][62]. There are also important confusions between neutral and sadness. Getting back to Fig. 15 on overall distribution of ratings, these confusions can be explained partly by the fact that intense anger and intense joy are much overlapped around value 2 on the arousal axis whereas neutral & intense sad are much overlapped about 0 on the arousal axis as well.

For comparison purpose, Fig. 18 gives the confusion matrix from DES. Similar confusions between anger and joy, and between sadness and neutral, can be observed on this widely used public dataset.

This phenomenon of surprising confusions suggests that the baseline feature set from INTERSPEECH 2013 is still not discriminative enough despite of its large scale. Indeed, using the proximal labels from the 7 raters, we can estimate a human emotion recognition rate on the proposed MES-P that we have reported in Table XIX. It simply averages the emotion recognition accuracies over the 7 raters and displays a 90.63% accuracy. The gap between the automatic classification and human perception in problem A (67.07% vs 90.63%) clearly indicate that novel and better discriminative features are required for better machine-based vocal emotion recognition.

*2) Problem B - classification on 7 emotional states:* In this case, the moderate and intense state of each emotion family are regarded as different emotion classes, leading to 7 classes.

The recognition accuracies are shown in the second row in Table XIX. As can be seen in Table XIX, due to the increase of number of classes, the various accuracies, *i.e.*, D-D, D-P, P-D, P-P, record a decrease of more than 20 points in comparison to problem *A*. However, for a classification problem with 7 classes, the chance level is 1/7, which is approximately 14.29%, a recognition accuracy around 45% in this leave-one-speaker-out protocol is still much over the chance level.

Fig. 19 shows the confusion matrices of problem *B*. As can be seen in this figure, the 2 predictive models, namely D-X and P-X, using distal labels and proximal labels for train-ing, respectively, display similar recognition accuracies. The predictive model D-X using distal labels for training displays much better recognition accuracy for emotional state intense joy (J-I) than the predictive model P-X using proximal labels for training. Conversely, the predictive model P-X exhibits better performance on neutral than the predictive model D-X. Similar confusion patterns can be observed as in problem A, *e.g*., intense anger (A-I) much confused with intense joy (J-I). But the addition of moderate emotional states as novel classes has also introduced novel confusion patterns, *e.g*., intense sadness (S-I) much confused with moderate sadness (S-M), neutral (N) with moderate sadness (S-M) or moderate anger (A-M) with neutral (N).

One natural question which arises here is whether the discrimination of all the 7 classes is feasible. What is the human performance facing the same problem. Using the proximal labels from the 7 raters in comparison with the distal labels, we have evaluated their average recognition accuracy over the 7 classes and found an accuracy of 86.54% (see second row in Table XIX). This accuracy is remarkable in comparison with the baseline performance whose average accuracy is only around 45%. In comparison with problem *A* where only 4 classes with neutral and intense emotional states are considered, the increase of number of classes due to moderate emotional states has only led to a decrease of 4 points in recognition accuracy for humans!

Furthermore, in comparing the confusion matrices in Fig. 19 with Table. XIV on speakers' expression ability or Table. XVI on raters' emotion sensitivity, it can be seen that recognition accuracies on different emotional states are significantly different between the baseline and humans. The worst recognized state by the baseline, neutral, with a recognition rate as low as around 33%, is the rather well expressed (92.74%) by speakers and well recognized (95.83%) by raters, without high confusions with moderate emotions as observed in the baseline. This ability of distinguishing emotion intensity by humans clearly contrasts with the high confusions introduced in the baseline due to the addition of moderate emotional states. For the 3 emotional families other than neutral, we can see the sequence from the best recognized to worst recognized is anger > joy > sadness. This is similar to the pattern in human perception (proximal labels) and speaker expression ability in Table. XVI and Table. XIV, respectively, but with much lower accuracy rates.

TABLE XVIII
COMPOSITION OF BASELINE FEATURE SET OF INTERSPEECH 2013. SOME OF THE FUNCTIONS ONLY APPLY TO SPECIFIC PARTS OF SPEECH, *I. E.* VOICED OR UNVOICED

| LLDs | Functions applied to LLDs and ΔLLDs | Functions applied to LLDs only |
|---|---|---|
| Sum of auditory spectrum (loudness) | quartiles 1 to 3, 3 inter-quartile ranges | mean of peak distances |
| Sum of RASTA-style filtered auditory spectrum | 1% percentile (≈ min), 99% percentile (≈ max) | standard deviation of peak distances |
| RMS Energy | position of min/max | mean value of peaks |
| Zero-Crossing Rate | percentile range 1% - 99% | mean value of peaks - arithmetic mean |
| RASTA-style auditory spectrum, bands 1-26 (0 - 8 kHz) | arithmetic mean, root quadratic mean | mean/std.dev. of rising/falling slopes |
| MFCC 114 | contour centroid, flatness | mean/std.dev. of inter maxima distances |
| Spectral energy 250 - 650Hz, 1k - 4kHz | standard deviation, skewness, kurtosis | amplitude mean of maxima /minima |
| Spectral Roll Off Point 0.25, 0.50, 0.75, 0.90 | rel. duration LLD is above/below 25/50/75/90% range | amplitude range of maxima |
| Spectral Flux, Entropy, Variance, Skewness, Kurtosis | rel. duration LLD is rising/falling | linear regression slope, offset, quadratic error |
| Slope, Psychoacoustic Sharpness, Harmonicity | rel. duration LLD has positive / negative curvature | quadratic regression a, b, offset, quadratic error |
| F0 by SHS + Viterbi smoothing, Probability of voicing | gain of linear prediction (LP), LP Coefficients 1 - 5 | percentage of non-zero frames |
| logarithmic HNR, Jitter (local, delta), Shimmer (local) | mean, max, min, std. dev. of segment length | |

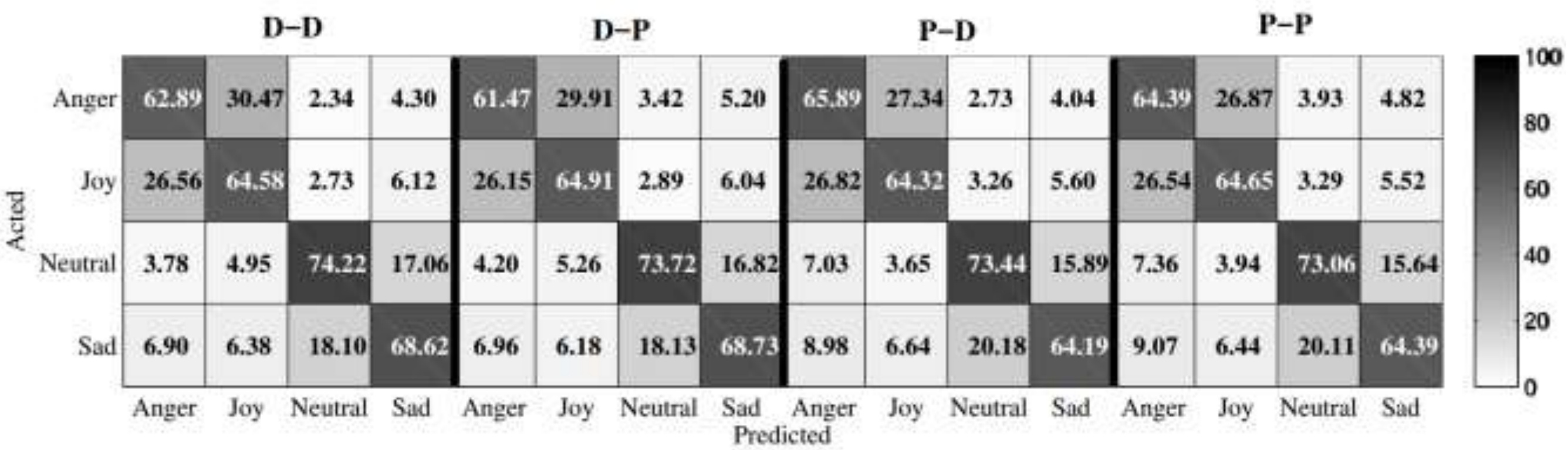

Fig. 17. Confusion matrices in problem A (%) on predictive model D-X and P-X, X designating D or P. D and P denote distal and proximal labels. Predictive model D-X (P-X, resp.) is trained using distal (P-X, resp.) labels and then evaluated using distal and proximal labels, leading to accuracies D-D and D-P (P-D and P-P, resp.).

TABLE XVIII
CORRECT CLASSIFICATION RATES (%) IN EACH PROBLEM

| Problem Code | D-D | D-P | P-D | P-P | Average | Human |
|---|---|---|---|---|---|---|
| A | 67.58 | 67.15 | 66.96 | 66.6 | 67.07 | 90.63 |
| B | 46.48 | 45.33 | 45.03 | 44.38 | 45.31 | 86.54 |

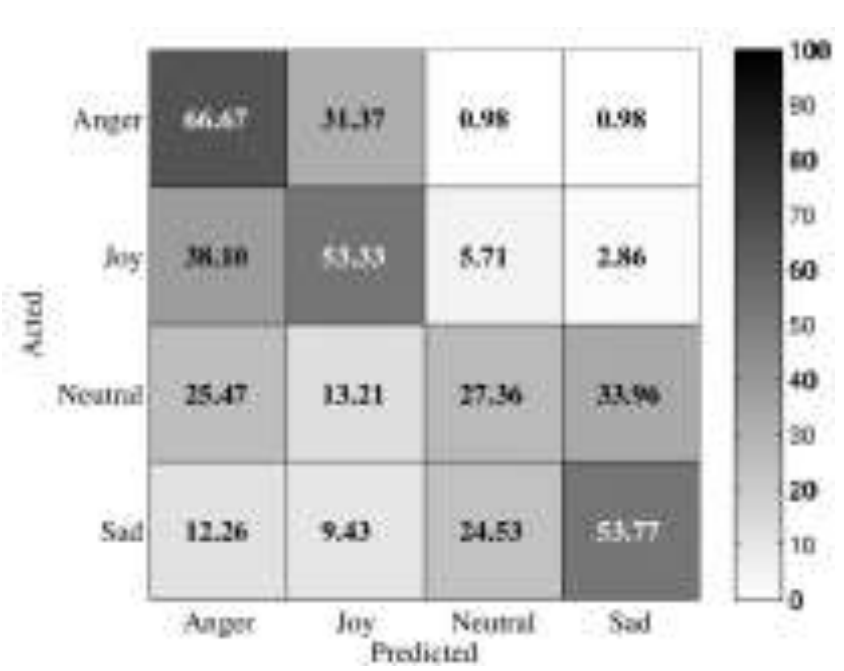

Fig. 18. Confusion Matrix on DES (%)

All these evidence suggest that the feature set used in the baseline is clearly not discriminative enough and better features are necessary.

## VII. CONCLUSION

An emotional speech dataset on Mandarin Chinese, namely MES-P, is developed and evaluated in this work. It includes three common emotion families, namely, joy, anger, and sadness, along with a neutral state, and features unique properties to get the proposed dataset closer to real-life vocal emotion communications. First, accounting for the fact that moderate emotions dominate in real-life vocal communications, MES-P investigates the intensity of emotions and distinguishes moderate mode from the intense mode for each emotion family. Then, to enable studies of tonal patterns in emotional speech in Mandarin Chinese, the proposed MES-P dataset displays a same distribution of the four Mandarin tones as in real-life communications and covers all types of vowels and consonants. Finally, we have also introduced in MES-P, for the first time to the best of our knowledge, *distal* and *proximal* labels which distinguishes, intended emotions that speakers encode into speech samples, from perceived emotions that listeners decode from pronounced vocal sentences. This distinction is of capital importance as it enables to characterize the level of vocal emotion intelligence of human beings, *i.e.*, emotion expression ability of speakers (How well emotions are encoded into vocal speech?) and vocal emotion sensitivity from listeners (How well a listener is able to understand the emotion encoded into vocal speech?), and quantify a human emotion recognition rate using proximal labels as human emotion predictions against distal labels as ground truth. The last but not the least is that we have provided a baseline which shows that much progress needs to be made so that machine-based vocal emotion recognition could compete with the human performance.

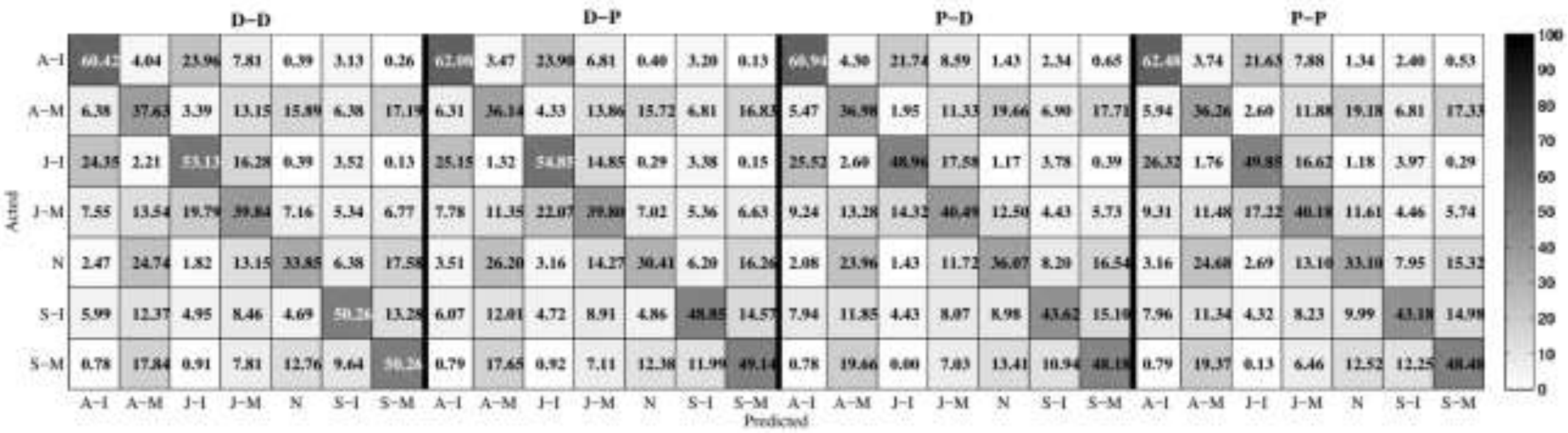

Fig. 19. Confusion matrices in problem B (%) on predictive model D-X and P-X, X designating D or P. D and P denote distal and proximal labels. Predictive model D-X (P-X, resp.) is trained using distal (P-X, resp.) labels and then evaluated using distal and proximal labels, leading to accuracies D-D and D-P (P-D and P-P, resp.). Letters A, J, S, N refer to emotions as anger, joy, sadness, neutral, and letters I and M refer to intense and moderate emotions respectively.

Our future work includes collection of emotional speech samples in uncontrolled environments with speakers and raters of varied ages, fine-grained emotion analysis, the impact of gender in emotion expression and recognition as well as better machine-based emotion recognition algorithms.

## REFERENCES

[1] R. Picard, “Affective Computing”, The MIT Press, 1997.
[2] Y. Baveye, C. Chamaret, E. Dellandréa, L. Chen, “Affective Video Content Analysis: A Multidisciplinary Insight”, IEEE Transactions on Affective Computing (Vol.PP, Issue:99), ISSN:1949-3045, DOI: 10.1109/TAFFC.2017.2661284, January 30, 2017
[3] O. Martin, I. Kotsia, B. Macq, et al. “The eNTERFACE’05 Audio-Visual Emotion Database”, International Conference on Data Engineering Workshops, 2006, pp. 8-15.
[4] M. Grimm, K. Kroschel and S. Narayanan, “The Vera am Mittag German audio-visual emotional speech database”, in Proc. ICME, Hannover, 2008, pp. 865-868.
[5] C. Busso, M. Bulut, C.C. Lee, et al. “IEMOCAP: interactive emotional dyadic motion capture database”, Language Resources and Evaluation, 2008, vol. 42, no. 4, pp. 335-359.
[6] G. McKeown, M. Valstar, R. Cowie, et al. “The SEMAINE corpus of emotionally coloured character interactions”, In Proc. ICME, Singapore, 2010, pp. 1079-1084
[7] G. McKeown, M. Valstar, R. Cowie, M. Pantic and M. Schroder, “The SEMAINE Database: Annotated Multimodal Records of Emotionally Colored Conversations between a Person and a Limited Agent”, in IEEE Transactions on Affective Computing, vol. 3, no. 1, pp. 5-17, Jan.-March 2012.
[8] H. Cao, D. G. Cooper, M. K. Keutmann, R. C. Gur, A. Nenkova and R. Verma, “CREMA-D: Crowd-Sourced Emotional Multimodal Actors Dataset”, in IEEE Transactions on Affective Computing, vol. 5, no. 4, pp. 377-390, Oct.-Dec. 1 2014.
[9] Y. Baveye, E. Dellandréa, C. Chamaret, L. Chen. “LIRIS-ACCEDE: A Video Database for Affective Content Analysis”, IEEE Transactions on Affective Computing, vol.6(1), pp:43-55, ISSN:1949-3045. DOI: http://dx.doi.org/10.1109/TAFFC.2015.2396531. 27/01/2015
[10] S. Jing, X. Mao, L. Chen, et al. “Annotations and consistency detection for Chinese dual-mode emotional speech database”, Journal of Beijing University of Aeronautics and Astronautics (in Chinese), 2015, vol. 41, no. 10, pp. 1925-1934.
[11] D. Ververidis, C. Kotropoulos, “Emotional speech recognition: Resources, features, and methods”, Speech Communication, 2006, vol.48, No.9, pp 1162-1181.
[12] K. R. Scherer, “Vocal communication of emotion: A review of research paradigms”, Speech Communication vol. 40, pp. 227-256, 2003.
[13] E. Brunswik, “Perception and the Representative Design of Psychological Experiments”, University of California Press, Berkeley, 1956.
[14] K.R. Hammond, T.R. Stewart, “The Essential Brunswik: Beginnings, Explications, Applications”, Oxford University Press, 2001, New York.
[15] L. Devillers, L. Vidrascu, L. Lamel, “Challenges in real-life emotion annotation and machine learning based detection”, Neural Networks, 2005, vol 18, pp. 407-422.
[16] P. Ekman, E.R. Sorenson, and W.V. Friesen, “Pan-cultural elements in facial displays of emotions”, Science, 1969, 164, pp. 86-88.
[17] P. Ekman, “An argument for basic emotions”, Cognition and Emotion, 1992, vol 6, no. 3/4, pp. 169200.
[18] W. Wundt, “Grundzuge der physiologischen Psychologie (Fundamentals of physiological psychology)”, originate published in 1874, 5th edition, 1905, Engelmann, Leipzig
[19] K.R. Scherer, “Psychological models of emotion”, The Neuropsychology of Emotion, Oxford University Press, Oxford/New York, 2000, pp. 137 - 162.
[20] J. Hansen, S. Bou-Ghazale, “Getting Started with SUSAS: A Speech Under Simulated and Actual Stress Database”, European Conference on Speech Communication & Technology, Eurospeech 1997, Rhodes, Greece, 1997: 1743-1745.
[21] I. S. Engberg, A.V. Hansen, O. Andersen, et al. “Design, Recording and Verification of a Danish Emotional Speech Database”, European

Conference on Speech Communication & Technology, 1997, pp. 1695-1698.
[22] E. Douglas-Cowie, R. Cowie, M. Schrder, “A new emotion database: considerations, sources and scope”, In Proc. ISCA, Belfast, UK, 2000, pp. 39-44.
[23] N. Campbell, “Databases of emotional speech”, In ITRW on Speech and Emotion, Newcastle, Northern Ireland, UK, 2000, pp. 34-38.
[24] F. Burkhardt. A. Paeschke, M. Rolfes, “A Database of German Emotional Speech”, in Proc2005 INTERSPEECH, Lisbon, Portugal, 2005, pp. 517-1520.
[25] D. Ververidis, C. Kotropoulos. “A state of the art review on emotional speech database”, in Proc 1st Rich Media Conference. Lausanne, Switzerland, 2003, pp. 109-119.
[26] R. Cowie, R. Cornelius, “Describing the emotional states that are expressed in speech”, Speech Communication, 2003, vol.40, pp. 5-32.
[27] S. Steidl, “Automatic classification of emotion-related user states in spontaneous children’s speech”, Ph.D. thesis, Berlin: Logos Verlag, 2009.
[28] B. Schuller, S. Steidl, A. Batliner. “The INTERSPEECH 2009 emotion challenge”, Interspeech, 2009, pp. 312-315.
[29] C¸ . Oflazoglu and S. Yildirim, “Turkish emotional speech database”, 2011 IEEE 19th Signal Processing and Communications Applications Conference (SIU), Antalya, 2011, pp. 1153-1156.
[30] I. Sneddon, M. McRorie, G. McKeown, and J. Hanratty, “The Belfast induced natural emotion database”, IEEE Trans. Affect. Comput., vol. 3, no. 1, pp. 3241, Jan.Mar. 2012.
[31] G. Costantini, I. Iadarola, A. Paoloni, M. Todisco, “EMOVO CORPUS: an Italian Emotional Speech Database,” in Proceedings of the 9th LREC, Reykjavik, Iceland, May 26-31, 2014, pp. 3501-3504.
[32] C. Busso, S. Parthasarathy, A. Burmania, M. AbdelWahab, N. Sadoughi and E. M. Provost, “MSP-IMPROV: An Acted Corpus of Dyadic Interactions to Study Emotion Perception”, in IEEE Transactions on Affective Computing, vol. 8, no. 1, pp. 67-80, Jan.-March 1 2017.
[33] Y. Zong, W. Zheng, T. Zhang, et al. “Cross-Corpus Speech Emo-tion Recognition Based on Domain-Adaptive Least-Squares Regression”. IEEE Signal Processing Letters, 2016, vol. 23, no. 5, pp. 585-589.
[34] Z. Zhang, E. Coutinho, J. Deng, et al. “Cooperative Learning and its Application to Emotion Recognition from Speech”, IEEE/ACM Transactions on Audio, Speech, and Language Processing, 2015, vol. 23, no. 1, pp. 115-126.
[35] J. Deng, Z. Zhang, F. Eyben, et al. “Autoenrater-based Unsupervised Domain Adaptation for Speech Emotion Recognition”, IEEE Signal Processing Letters, 2014, vol. 21, no. 9, pp. 1068-1072.
[36] K. Wang, N. An, B. Li, et al. “Speech Emotion Recognition Using Fourier Parameters”, IEEE Transactions on Affective Computing, 2015, vol. 6, no. 1, pp. 69-75.
[37] D. Bone, C. Lee, S. Narayanan, “Robust Unsupervised Arousal Rating: A Rule-Based Framework with Knowledge-Inspired Vocal Features”, IEEE Transactions on Affective Computing, 2014, vol. 5, no. 2, pp. 201-213.
[38] [38] Z. Xiao, D. Wu, X. Zhang, Z. Tao, “A Cross-Corpus Recognition of Emotional Speech”, in Proceedings of ISCID 2016, Hangzhou, Zhejiang, China, 2016, vol. 2, pp. 42-46.
[39] J. Tao, Fang Zheng, A. Li and Ya Li, “Advances in Chinese Natural Language Processing and Language resources”, 2009 Oriental COCOSDA International Conference on Speech Database and Assessments, Urumqi, 2009, pp. 13-18.
[40] M. Fox. “Zhou Youguang, Who Made Writing Chinese as Simple as ABC, Dies at 111”. The New York Times, 2017-01-14.
[41] “Pinyin celebrates 50th birthday”. Xinhua News Agency. 2008-02-11.
[42] “ISO 7098:1982 Documentation Romanization of Chinese”.
[43] B. Li, Z. Xiao, Y. Shen, Q. Zhou and Z. Tao, “Emotional speech conversion based on spectrum-prosody dual transformation”, 2012 IEEE 11th International Conference on Signal Processing, Beijing, 2012, pp. 531-535.
[44] Y, R, Chao, “A system of tone-letters”, Le Matre Phontique, 1930, vol. 45, pp. 2427.
[45] Character frequency table with 6763 characters, from the internet, https://wenku.baidu.com/view/2111ccfaaef8941ea76e051f.html
[46] https://en.wikipedia.org/wiki/GB 2312
[47] W. C. Hannas, “Asia’s Orthographic Dilemma”, University of Hawaii Press, 1997, p. 264.
[48] L.R. Derogatis, R.S Lipman, Covi, L. “SCL-90: An outpatient psychiatric rating scale-Preliminary Report”. Psychopharmacol. Bull. 1973, 9, 1328.
[49] L.R. Derogatis. “Symptom Checklist-90-Revised. in Handbook of psychiatric measures”. American Psychiatric Association 2000:pp.81-84.
[50] M. Holi, “Assessment of psychiatric symptoms using the SCL-90”. Doctoral Thesis, 2003, Helsinki: University of Helsinki.
[51] Technical sheet of AT2020USB, https://www.audio-technica.com.hk/templates/index/file/416_1_AT2020USB.pdf
[52] X. Zhao, J. Liu and K. Deng, “Assumptions behind intercoder reliability indices”, in Charles T. Salmon (ed.), Communication Yearbook 36, pp. 419-480. New York: Routledge, 2013
[53] J. Cohen, “A coefficient of agreement for nominal scales”, Educational and Psychological Measurement, 1960, vol. 20, no. 1, pp. 37-46.
[54] J. Cohen, “Nominal scale agreement with provision for scaled disagreement or partial credit”. Psychological Bulletin, 1968, vol. 70, pp. 213-220.
[55] E. Rogot, I.D. Goldberg, “A proposed index for measuring agreement in test-retest studies”. Journal of Chronic Diseases, 1966, vol. 19, no. 9, pp. 991-1006.
[56] Landis, J.R.; Koch, G.G. (1977). “The measurement of observer agreement for categorical data”. Biometrics. 33 (1): 159174.
[57] S. Hall, E. Frank, G. Holmes, B. Pfahringer, P. Reutemann, I. H. Witten, “The WEKA Data Mining Software: An Update”, SIGKDD Explorations, 2009, Volume 11, Issue 1.
[58] B. Schuller, S. Steidl, A. Batliner, A. Vinciarelli, K. Scherer, R. Ringeval, M. Chetouani, F. Weninger, F. Eyben, E. Marchi, M. Mortillaro, H. Salamin, A. Polychroniou, F. Valente, S. Kim. “The INTERSPEECH 2013 computational paralinguistics challenge: Social signals, conflict, emotion, autism”. Proceedings of the Annual Conference of the International Speech Communication Association, INTERSPEECH, 2013. pp 148-152.
[59] B. Schuller,S. Steidl, A. Batliner, E. Bergelson, J. Krajewski, C. Janott, et al. “The INTERSPEECH 2017 computational paralinguistics challenge: Addressee, cold & snoring”. In Proceedings of Interspeech 2017, pp.3442-3446.
[60] F. Eyben, M. W ¨ ollmer, and B. Schuller, “openSMILE - The Munich Versatile and Fast Open-Source Audio Feature Extractor”, in Proc. ACM Multimedia. Florence, Italy: ACM, 2010, pp. 1459-1462.
[61] D. Ververidis, C. Kotropoulos,“Emotional Speech Classification Using Gaussian Mixture Models and the Sequential Floating Forward Selection Algorithm”, IEEE International Conference on Multimedia and Expo, 2005. ICME 2005, p1500-1503.
[62] Z. Xiao, “Recognition of Emotions in Audio Signals”, Ph.D. Thesis, Ecole Centrale de Lyon, 2008.